\documentclass{article}
\usepackage[utf8]{inputenc}

\usepackage{multirow}
\usepackage{graphicx}
\usepackage{amsmath}
\usepackage[english]{babel}

\begin{document}

\title{Simulations of Dynamical Gas–Dust Circumstellar Disks: \\
Going Beyond the Epstein Regime}

\author{O.~P. Stoyanovskaya, F.~A. Okladnikov, E.~I. Vorobyov, \\
Ya.~N. Pavlyuchenkov, V.~V. Akimkin \\ 
Lavrentyev Institute of Hydrodynamics,\\ Siberian Branch of the Russian Academy of Sciences, \\
Novosibirsk, Russia, \\
Institute of Computational Technology,\\ Siberian Branch of the Russian Academy of Sciences, \\Novosibirsk, Russia,\\ Novosibirks State University, Novosibirsk, Russia \\ 
Southern Federal University, Institute of Physics, Rostov-on-Don, Russia, \\ Department of Astrophysics, University of Vienna, Vienna, Austria \\
Institute of Astronomy, Russian Academy of Sciences, \\ Moscow, Russia}

\maketitle

\begin{abstract}

In circumstellar disks, the size of dust particles varies from submicron to several centimeters, while planetesimals have sizes of hundreds of kilometers. Therefore, various regimes for the aerodynamic drag between solid bodies and gas can be realized in these disks, depending on the grain sizes and velocities: Epstein, Stokes, and Newton, as well as transitional regimes between them. This means that simulations of the dynamics of gas–dust disks require the use of a drag coefficient that is applicable for a wide range for sizes and velocities for the bodies. Furthermore, the need to compute the dynamics of bodies of different sizes in the same way imposes high demands on the numerical method used to find the solution. For example, in the Epstein and Stokes regimes, the force of friction depends linearly on the relative velocity between the gas and bodies, while this dependence is non-linear in the transitional and Newton regimes. On the other hand, for small bodies moving in the Epstein regime, the time required to establish the constant relative velocity between the gas and bodies can be much less than the dynamical time scale for the problem—the time for the rotation of the disk about the central body. In addition, the dust may be concentrated in individual regions of the disk, making it necessary to take into account the transfer of momentum between the dust and gas. It is shown that, for a system of equations for gas and monodisperse dust, a semi-implicit first-order approximation scheme in time in which the interphase interaction is calculated implicitly, while other forces, such as the pressure gradient and gravity are calculated explicitly, is suitable for stiff problems with intense interphase interactions and for computations of the drag in non-linear regimes. The piece-wise drag coefficient widely used in astrophysical simulations has a discontinuity at some values of the Mach and Knudsen numbers that are realized in a circumstellar disk. A continuous drag coefficient is presented, which corresponds to experimental dependences obtained for various drag regimes.

\end{abstract}


\section{INTRODUCTION}
Simulations of the dynamics of circumstellar gas–dust disks are necessary to improve our understanding of mechanisms for the formation of planets, and have been carried out in many studies, such as \cite{BaiStone2010ApJS,ZhuDust,ChaNayakshinDust2011,RiceEtAl2004,FranceDustCode,Pignatale2016,VorobyovEtAl2017,1MNRAS,DemidovaGrininDust2017}. Thus far, models for the dynamics of gas–dust disks have been developed in which the gas is taken to be a carrier phase, and solid particles to be a dispersed phase. In this case, the dynamics of a dust cloud can be described like the dynamics of a continuous medium (the basis for applying this approximation is presented in Appendix A). If we suppose that the solid phase is represented by particles of a single size at each point in space, the continuity equation and equation of motion of the gas and dust in the disk will have the form
\begin{equation}
\label{eq:gas}
\displaystyle\frac{\partial \rho_{\rm g}}{\partial t}+\nabla (\rho_{\rm g} v)=S_{\rm g},\ \ \ 
 \rho_{\rm g} \left[\displaystyle\frac{\partial v}{\partial t}+(v \cdot \nabla) v \right]=-\nabla p+ \rho_{\rm g} g - f_{\rm D}+f_{\rm g},
\end{equation}
\begin{equation}
\label{eq:dust}
\displaystyle\frac{\partial \rho_{\rm d}}{\partial t}+\nabla (\rho_{\rm d} u)=S_{\rm d},\ \ \ 
\rho_{\rm d} \left[\displaystyle\frac{\partial u}{\partial t}+(u \cdot \nabla) u \right]=\rho_{\rm d} g + f_{\rm D}+f_{\rm d},
\end{equation}
where $\rho_{\textrm{g}}$ and $\rho_{\textrm{d}}$ are the volume densities of the gas and dust, $v$ and $u$ the velocities of the gas and dust, $p$ the gas pressure, $g$ the acceleration due to the self-gravitation of the gas and dust, as well as the gravity of the central star, $S_{\textrm{g}},S_{\textrm{d}}$ the sources and sinks for the gas and dust, $f_{\textrm{g}},f_{\textrm{d}}$ the forces acting on the gas and dust, apart from the pressure force, gravitation, and drag, $f_{\textrm{D}}$ the drag force per unit volume of the medium, 
\begin{equation}
    f_{\rm D}=n_{\rm d}F_{\rm D},
\end{equation}
where $n_{\rm d}$ is the volume number density of dust particles, $F_{\rm D}$ the force of friction per particle,
\begin{equation}
F_{\rm D}=\frac{1}{2} C_{\rm D} s \rho_{\rm g} \|v-u\|(v-u),\\ 
\label{eq:DragForce}
\end{equation}
where $s$ is the area of the “frontal” cross section of a particle (for example, for a sphere of radius ), and $C_{\rm D}$ is the dimensionless drag coefficient, which is a function of two dimensionless quantities – the Mach number,
\begin{equation}
    \textrm{Ma}=\frac{\|v-u\|}{c_s}
\end{equation}
and the Knudsen number,
\begin{equation}
    \textrm{Kn}=\frac{\lambda}{a},
\end{equation}   
Here, $\lambda$ is the mean free path of a gas molecule $c_s$ is a sound speed in pure gas.

The dust and solid bodies in the circumstellar disk comprise objects with various sizes, from submicron dust particles to planetesimals up to hundreds of kilometers. Large and small bodies interact differently with the gas. Small dust particles intensively exchange momentum with the gas, and their steady-state or terminal velocities relative to the gas are small. Large bodies can have velocities that are appreciably different from the gas velocity. This means that $\rm{Ma}$, $\rm{Kn}$ and $||f_{\rm D}||$  in the disk vary by orders of magnitude. Consequently, for numerical simulations of the dynamics of the two-phase medium in the disk, it is necessary to use a coefficient of friction that is applicable over a broad range of values of $\rm{Ma}$ and $\rm{Kn}$. Therefore, \ref{sec:DragDisc}  presents an analysis of various forms of drag coefficients and shows important differences between them and their advantages when applied in models of a disk. Furthermore, $||f_{\rm D}||$ in Eqs.~(\ref{eq:gas}) and (\ref{eq:dust}) appreciably exceeds the sum of the other forces for small bodies, but is comparable to the other forces for large bodies. In general, $f_{\rm D}$ depends non-linearly on the relative velocity between the gas and bodies. These factors impose high demands on  the numerical method used to solve the equations of motion (\ref{eq:gas}) and (2). Section \ref{sec:previos work} presents a review of this problematic situation and published results of studies on methods for the simulation of the dynamics of a gas–dust medium with intense interphase interactions. Section \ref{sec:scheme} presents a semi-implicit scheme that is first order in time, and Section \ref{sec:test} shows that this scheme satisfies the necessary requirements for its application in simulations of circumstellar disks. The conclusions are presented in Section \ref{sec:resume}. The applicability of the model to a continuous–medium description of the dynamics of the dust in the disk is justified in Appendix A, where alternative approaches are also presented. Appendix B describes the simple, one-dimensional model for a circumstellar disk used for our simulations.  

\section{DRAG COEFFICIENTS}
\label{sec:DragDisc}

\begin{figure}
    \centering
    \includegraphics[width=150mm]{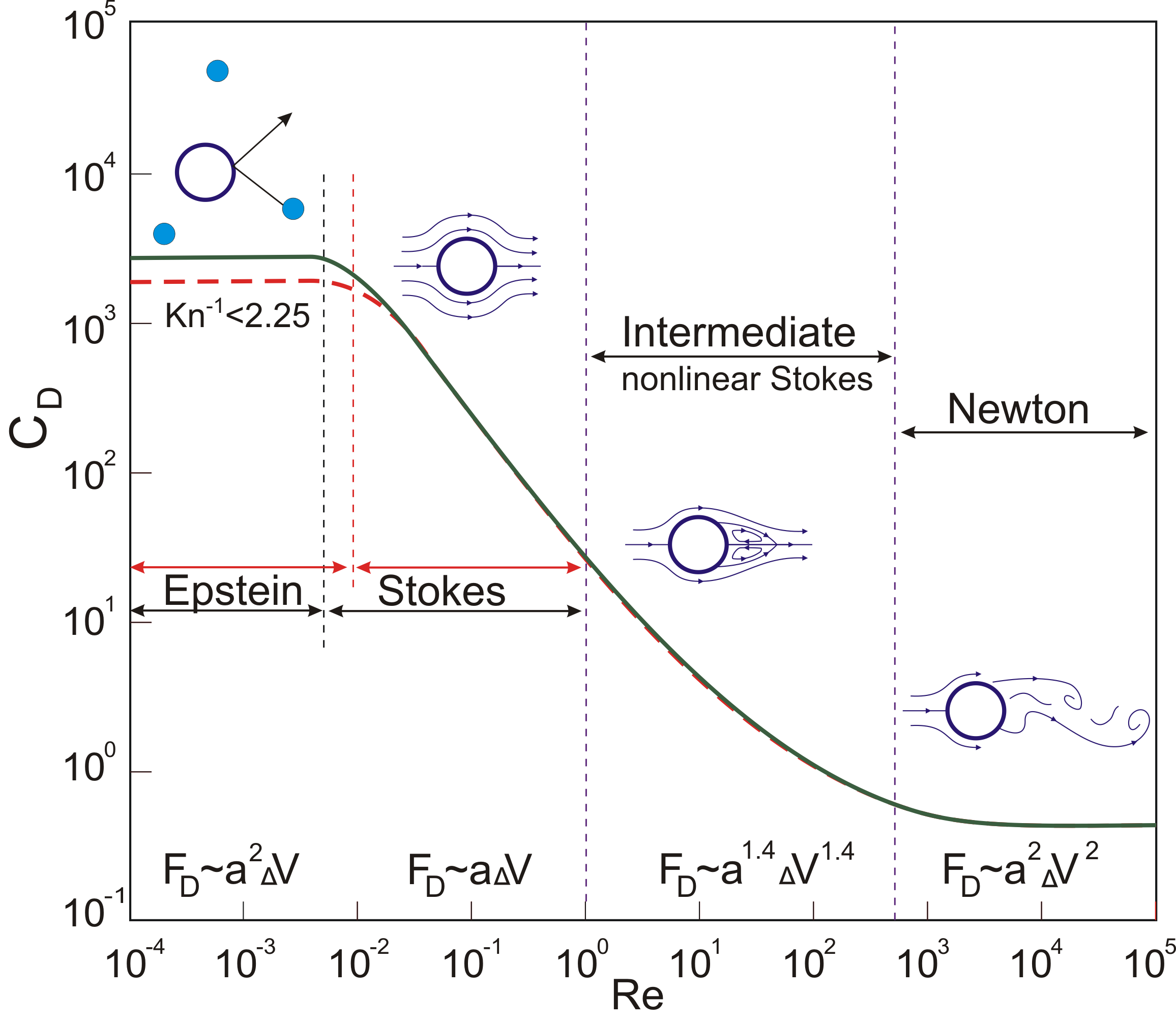}
    \caption{Drag coefficient as a function of the Reynolds and Mach numbers for the free-molecular flow regime (Epstein regime) and the case when the matter flows around a body like a continuous medium (the Stokes, transition, and Newton regimes). The boundary between free-molecular flow and continuous-medium flow is determined by the Knudsen number, and the boundaries of the continuous-medium flow regime by the Reynolds number. Stream lines around spherical particles are shown for the continuous-medium flow regimes. The functional dependence of the drag force $F_{\rm D}$ on the particle radius $a$ and relative velocity between the particles and gas $\Delta V=\|v-u\|$ are presented for all regimes. }
    \label{fig:CDRe}
\end{figure}

In the case of a compressible gas, the drag coefficient $C_{\rm D}$ in (\ref{eq:DragForce}) is a function of the two independent variables $\rm{Ma}$ and $\rm{Kn}$. The characteristic form of $C_{\rm D}$ is presented in Fig.~\ref{fig:CDRe}. In a regime where the gas flows around the bodies like a continuous medium, the drag coefficient is determined by the Reynolds number $\rm{Re}$ (see, for example, \cite{BagheriBonadonna2016}), which is derived from $\rm{Ma}$ and $\rm{Kn}$:
\begin{equation}
    \rm{Re}=4\frac{\rm{Ma}}{\rm Kn}.
\end{equation}
Particles with radii less than  $2.25\lambda$, interact with the gas in a free-molecular-flow regime, or Epstein regime \cite{Epstein1924}. The drag coefficients for such bodies do not depend on $\rm{Re}$, but do depend on $\rm{Ma}$. This dependence corresponds to the standard astrophysical coefficient proposed in \cite{Weidenschilling1977}:

\begin{equation}
\label{eq:CD4}
C_{\rm D}=
\left\{
	\begin{array}{lcl}
    \displaystyle
	\frac{8}{3}\rm{Ma}^{-1}, \quad \rm{Kn^{-1}}<\frac{9}{4} \quad \textrm{(Epstein regime; I),}\\[10pt]
    \displaystyle
    24\rm{Re}^{-1},\quad \rm{Re}<1 \quad \textrm{(Stokes regime; II),}\\[10pt]
    \displaystyle
    24\rm{Re}^{-0.6},\quad 1<\rm{Re}<800 \quad \textrm{(transition regime
or non-linear Stokes regime; III),}\\[10pt]   
    \displaystyle
    0.44,\quad \rm{Re}>800 \quad  \textrm{(Newton regime; IV).}
	\end{array}
    \right.
\end{equation}
\vspace{10pt}

An approximation for the Stokes regime, the transition regime (also known as the non-linear Stokes regime), and the Newton regime in (\ref{eq:CD4}) was proposed in \cite{ProbsteinFassio1970}, and was applied in \cite{Whipple1972} to the aerodynamics of solid bodies in a disk. This approximation was expanded to encompass the Epstein regime in \cite{Weidenschilling1977}. The coefficient~(\ref{eq:CD4}) is used in \cite{RiceEtAl2004,LaibePriceAstroDrag} and other studies in the framework of modern models for circumstellar disks. 

\begin{figure}
    \centering
    \includegraphics[width=110mm]{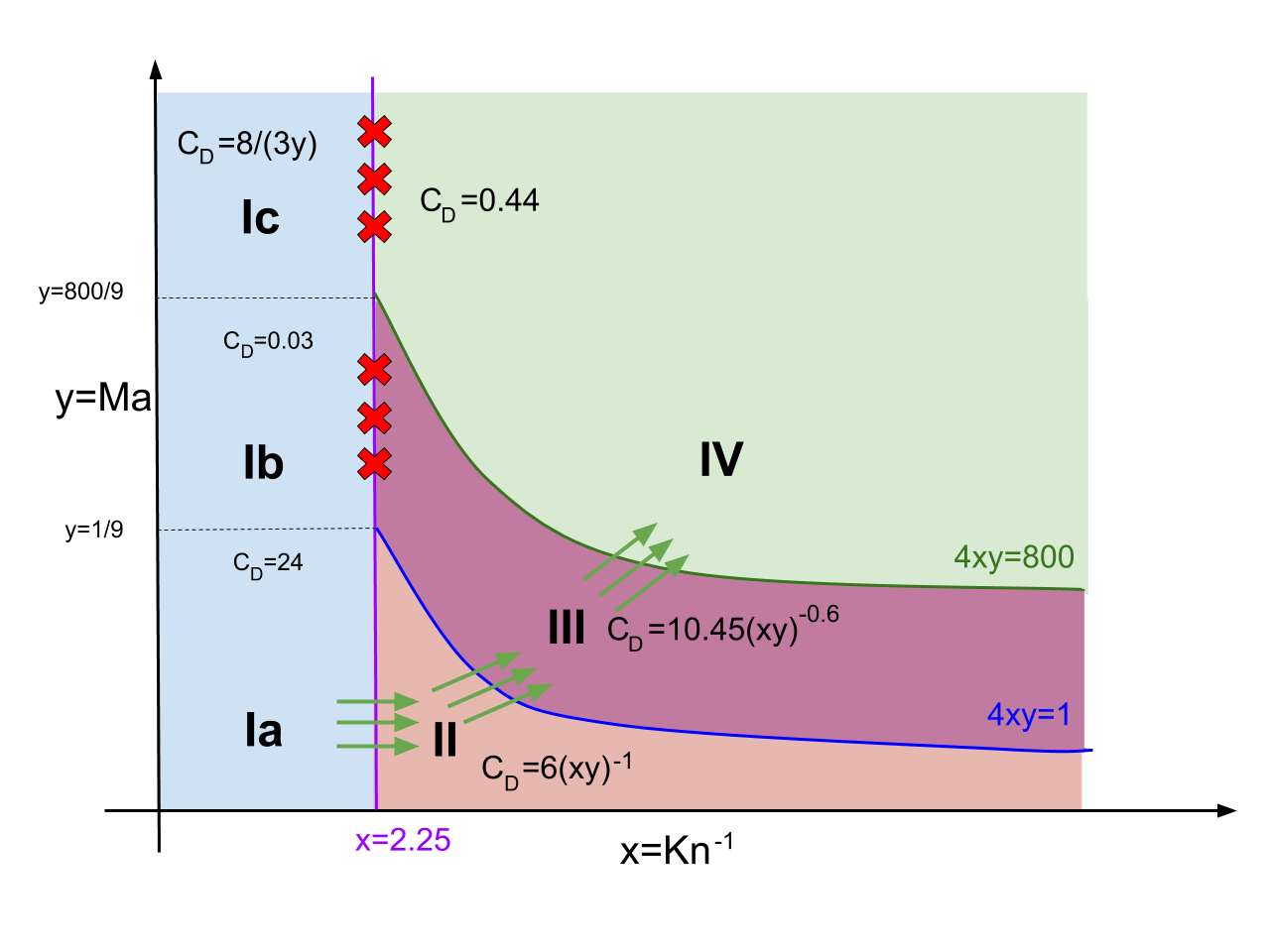}
    \caption{Regions of determination of the piecewise function $C_{\rm D}$ (\ref{eq:CD4}), corresponding to the various flow regimes. I—Epstein, II—Stokes, III—transition (non-linear Stokes), IV—Newton. The green arrows indicate the boundaries at which a function is continuous, and the red x’s boundaries where a function is discontinuous.}
    \label{fig:CD4pieces}
\end{figure}
 
Figure~\ref{fig:CD4pieces} presents the regions I–IV within which the drag coefficient $C_{\rm D}$ (as a function of $\rm{Kn^{-1}}$ and $\rm{Ma}$) is continuously differentiable. At the boundaries I–II, II–III, and III–IV, $C_{\rm D}$ is continuous, whereas this coefficient is discontinuous at the boundaries I–III and I–IV. The presence of a discontinuity is a serious inadequacy for simulations, since this can lead to the appearance of artificial singularities in the solution. We convinced ourselves that the conditions for which a particle makes the transition from the Epstein regime I to the transition regime (non-linear Stokes regime) III, which is accompanied by a discontinuity of the coefficient of friction (\ref{eq:CD4}). is indeed realized in simulations of the dynamics of a disk. This conclusion was based on the simulations of the dynamics of a viscous, self-gravitating, gaseous disk with a dust component using the FEOSAD model, which is described in detail in \cite{VorobyovEtAl2017}. The disk model used differs from the basic model presented in \cite{VorobyovEtAl2017}in the additional consideration of the influence of the dust dynamics on the gas dynamics, and also in the absence of the restriction on the size of the dust $a<2.25 \lambda$, which artificially maintains the particles in the Epstein regime. We used drag coefficient (\ref{eq:CD4}) and the numerical scheme from Section \ref{sec:scheme} for our simulations. We used a constant value for the Shakura–Syunyaev viscosity over the disk, $\alpha=10^{-4}$, and the fragmentation velocity $v_{\rm frag}=30$~m~s$^{-1}$. This Shakura–Syunyaev viscosity corresponds to a disk with suppressed magnetorotational instability, in which the transport of mass and angular momentum occurs mainly via gravitational torques in a non-axially symmetric disk. A fragmentation velocity $v_{\rm frag}$ determines the maximum particle size. The fragmentation velocity chosen corresponds to an enhanced probability of adhesion, as is realized when the particles are covered in a thick layer of ice.

\begin{figure}
    \centering
    \includegraphics[width=100mm]{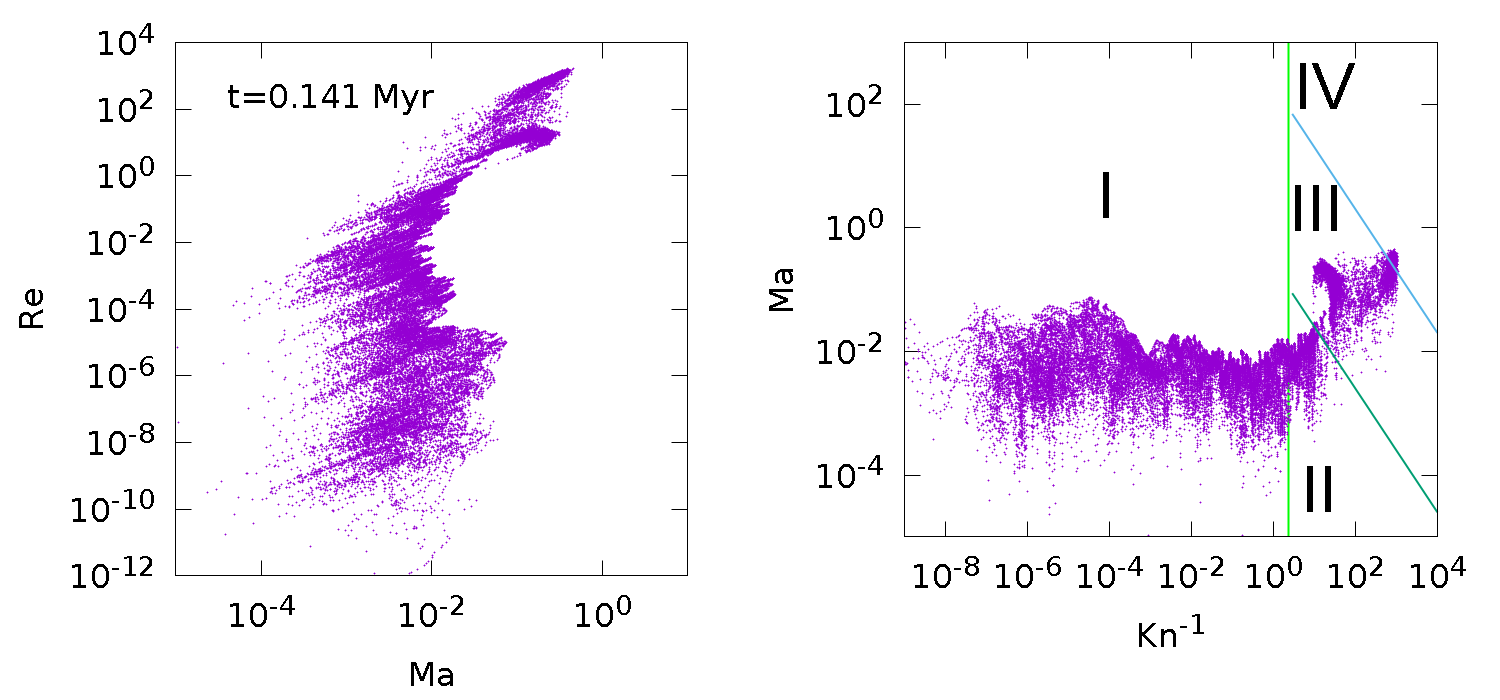}
    \includegraphics[width=50mm]{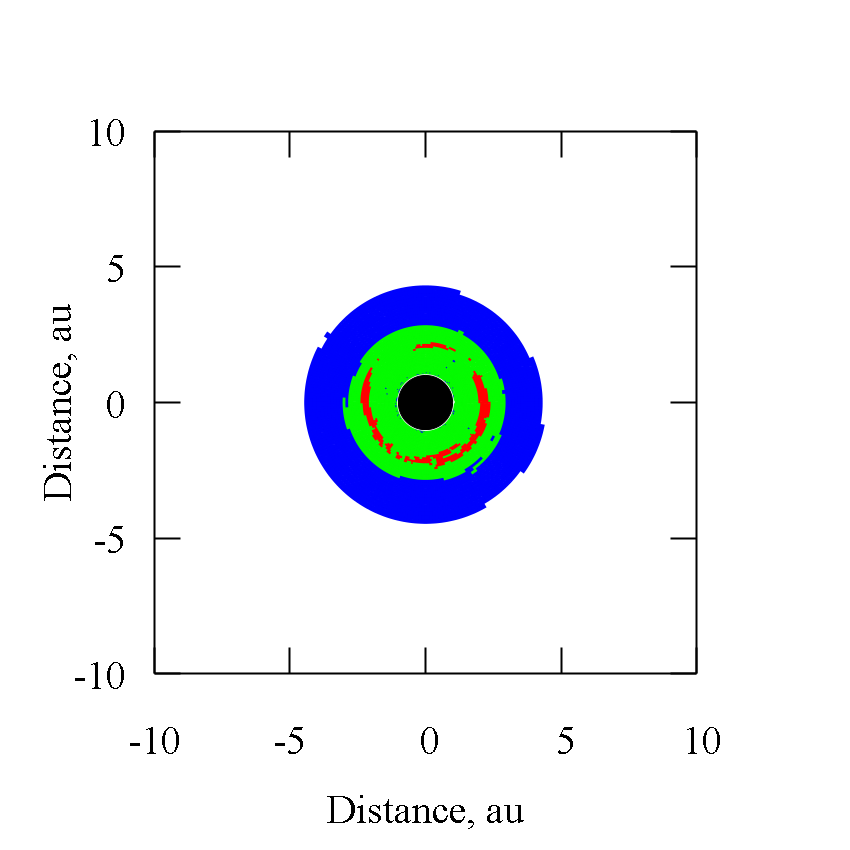}
    \includegraphics[width=100mm]{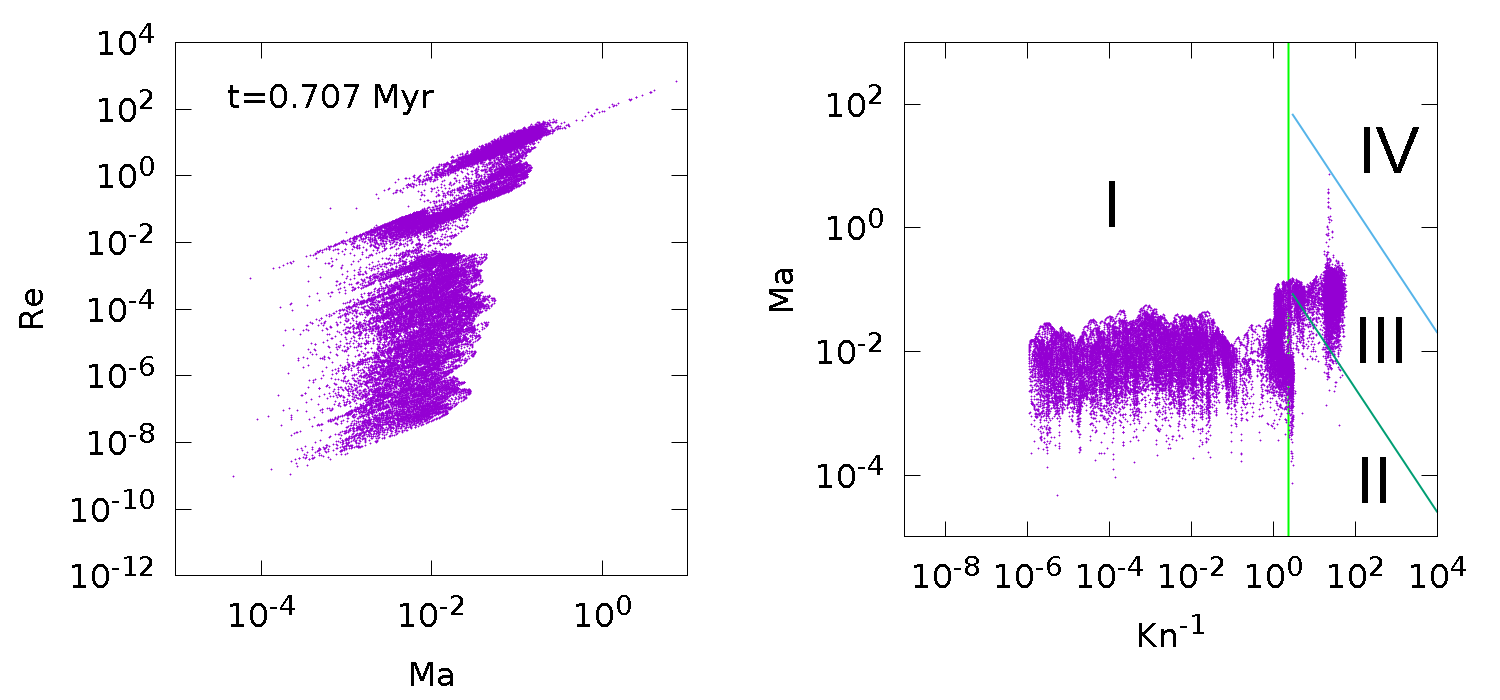}
    \includegraphics[width=50mm]{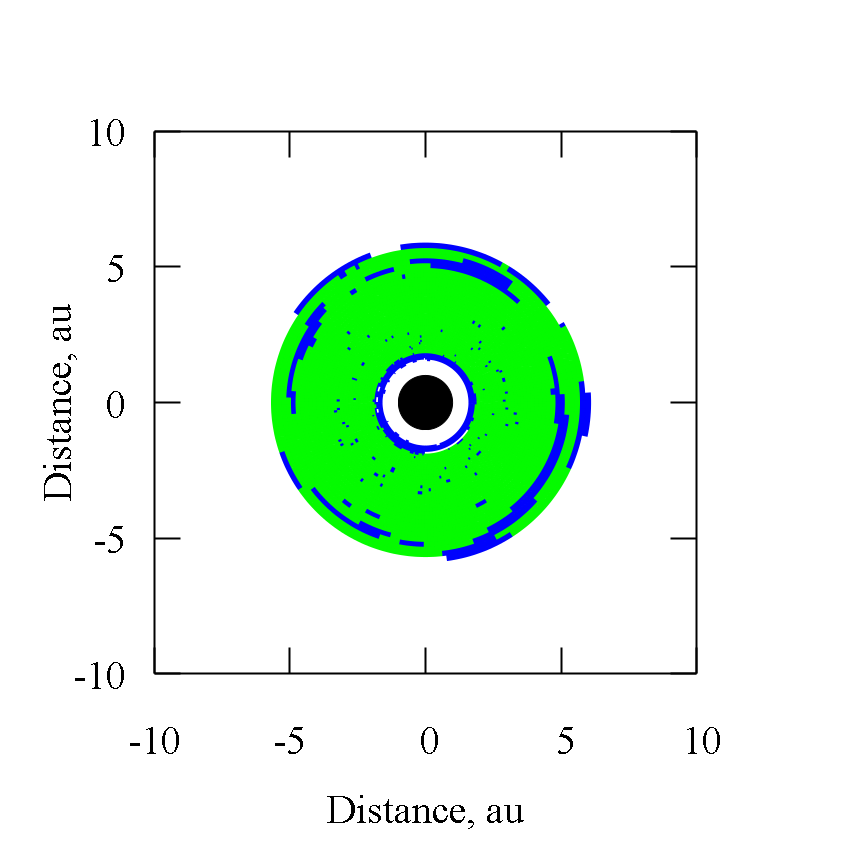}
    \caption{Values of $\rm{Ma}$ and $\rm{Re}$ (left column) and of $\rm{Ma}$ and $\rm{Kn}^{-1}$ (central column) for dust grains in all cells of circumstellar disks with ages of 141 000 years (upper row) and 707000 years (lower row). The various drag regimes from (\ref{eq:CD4}) are shown in the central panels by Roman numerals and in the right panels by different colors. I—Epstein (white), II—Stokes (blue), III—transition (non-linear Stokes) (green), IV—Newton (red). Black indicates drag cells of the disk. The standard astrophysical drag coefficient (\ref{eq:CD4}) is discontinuous at the boundary between the white and green region.}
    \label{fig:MaRe_St3}
\end{figure}

\begin{figure}
    \centering
    \includegraphics[width=80mm]{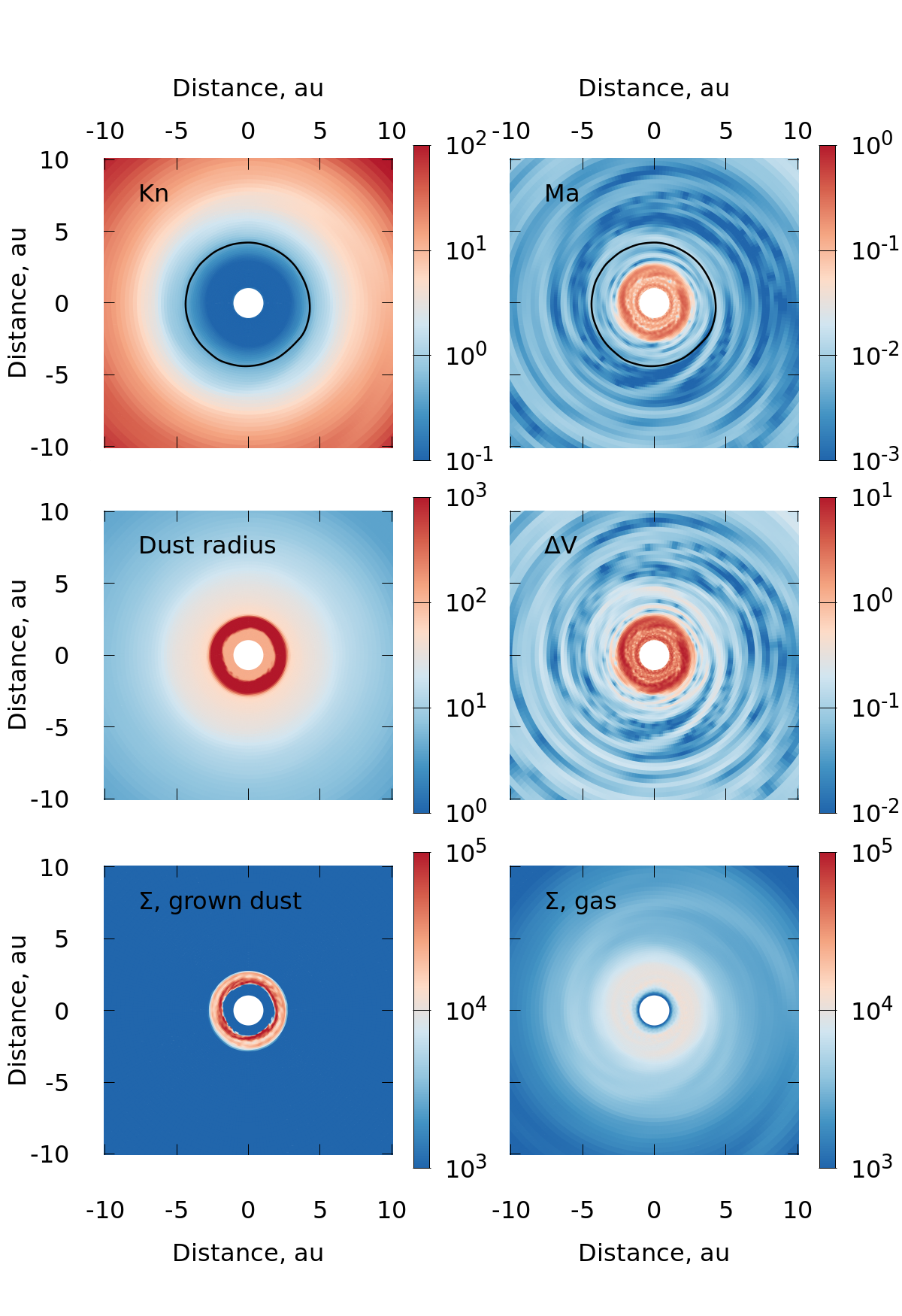}
    \includegraphics[width=80mm]{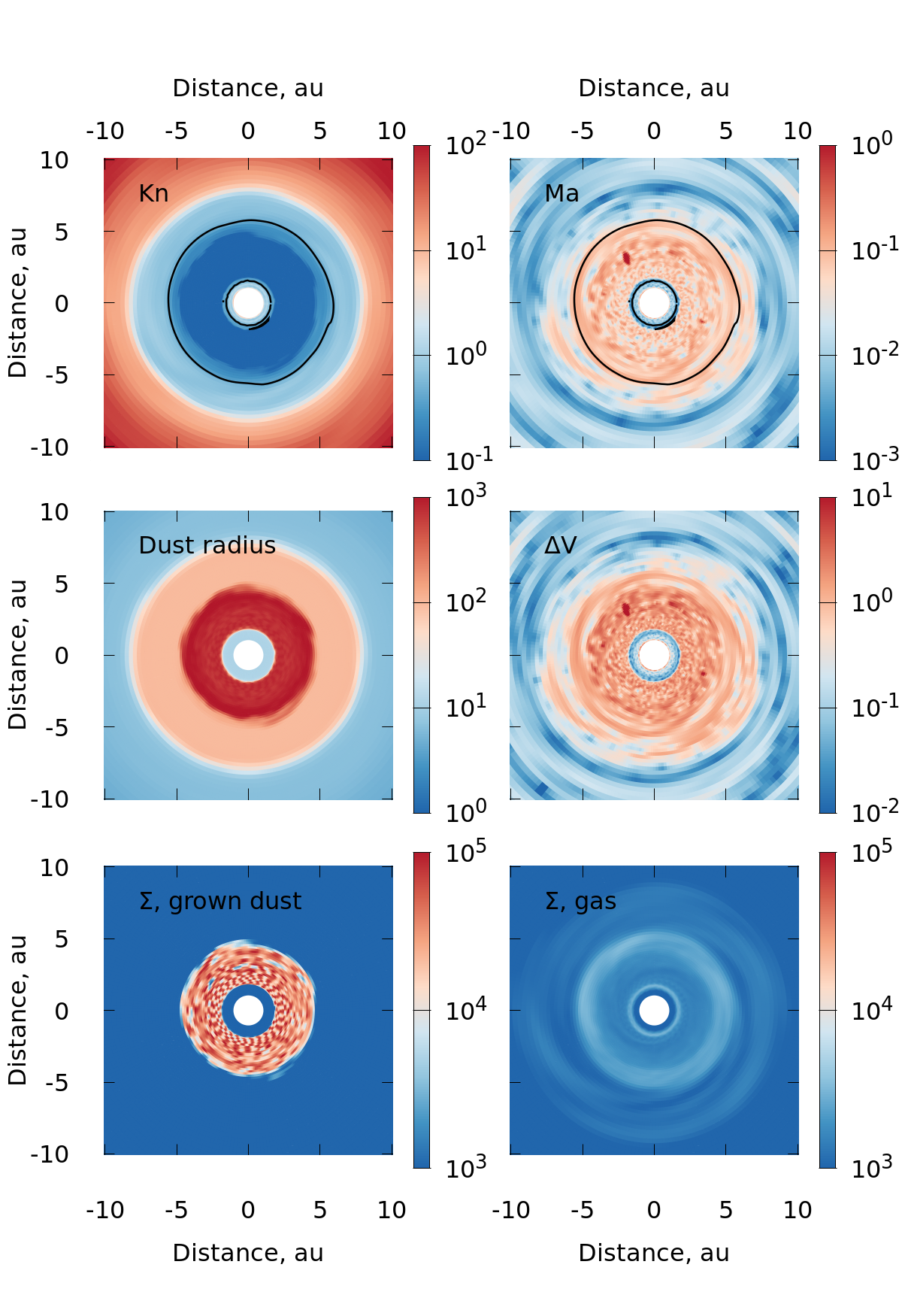}
    \caption{Distributions of physical quantities determining the drag regimes in the central regions of circumstellar disks with ages of 141000 years (left six panels) and 707000 years (right six panels). The upper panels present the Knudsen and Mach numbers and the lower panels present the surface densities of gas and grown dust in g cm$^{-2}$, The central panels show the maximum grain size $a$ in cm and the relative velocity between the gas and dust $\Delta V$ in km~s$^{-1}$. The black contours in the upper panels show the boundary at which $a=4 \lambda /9$. For a disk with an age of 141000 years, this passes through the region $\rm{Ma} \approx 0.01$, providing a continuous drag coefficient, while it passes through the region $\rm{Ma} \approx 1/9$, leading to a discontinuous drag coefficient for a disk with an age of 707000 years. }
    \label{fig:twoMoments}
\end{figure}

We computed the quantities $\rm{Kn}$, $\rm{Ma}$ and $\rm{Re}$ in each computational cell for the evolution of the disk over one million years. Figure \ref{fig:MaRe_St3} presents the set of values of these quantities in the disk in the $\rm{Ma}$, $\rm{Re}$ (left panels) and $\rm{Kn}^{-1}$, $\rm{Ma}$ (central panels) axis for ages of 141000 and 707000 years. The central panels show that the dust and gas interact in the free-molecular flow Epstein regime in the most part of the disk, however all four of the drag regimes indicated in (\ref{eq:CD4}) are realized in the disk. At a time of 707000 years (central panel in the second row), particles are present in the disk at the boundary of regions I–III, where the drag coefficient (\ref{eq:CD4}) undergoes a discontinuity. The right lower panel in Fig.\ref{fig:MaRe_St3} where different colours indicate different regions in the disk with different drag regimes, shows that this discontinuity occurs at a distance of 6 au from the star. The distribution of physical quantities determining the drag regime in the central region of the circumstellar disk at the same times are shown in Fig.~\ref{fig:twoMoments}. The black contour in the upper panels, which present the Knudsen and Mach numbers, shows the boundary where $a=4 \lambda /9$. In a disk with an age of 141000 years, this boundary passes through the region $\rm{Ma} \approx 0.01$, corresponding to a continuous coefficient of friction, while it passes through the region $\rm{Ma} \approx 1/9$ in a disk with an age of 707000 years, indicating the presence of a discontinuity in the drag coefficient.

Therefore, we investigated an alternative drag coefficient for cases when the medium flowing around the particles is compressible—the coefficient of Henderson \cite{Henderson1976}. The approximation of Henderson provides a good description of known experimental dependences for $\rm{Re}<3 \times 10^{5}$ and $\rm{Ma}<6$(details can be found in \cite{Henderson1976} and references therein). It takes into account the difference between the gas and particle temperatures. For subsonic flow regimes (${\rm Ma} \leq 1$) 

\begin{equation}
    C_{\rm D}=\displaystyle\frac{24}{\mathrm{Re}+S \left( 4.33+\displaystyle\frac{3.65-1.53 T_{\rm d}/ T_{\rm g}}{1+0.353 T_{\rm d}/T_{\rm g}}  \exp(-0.247 \frac{\rm{Re}}{S})\right)}+0.6S\left(1-\exp\left(-\displaystyle\frac{\rm Ma}{\rm Re}\right)  \right)
\end{equation}
\begin{equation}
\label{eq:HendersonSub}
    +\exp\left(-0.5\displaystyle\frac{\rm Ma}{\sqrt{\rm Re}}\right) \left( \displaystyle
    \frac{4.5+0.38 \left(0.03 {\rm Re}+0.48 \sqrt{\rm Re}\right)}{1+0.03 {\rm Re} + 0.48 \sqrt{\rm Re} } +0.1 {\rm Ma}^2+0.2 {\rm Ma}^8 \right)
\end{equation}
while for supersonic regimes (${\rm Ma} \geq 1.75$) 
\begin{equation}
\label{eq:HendersonSuper}
   C_{\rm D}=\displaystyle\frac{0.9+\displaystyle \frac{0.34}{\rm{Ma}^2}+1.86\sqrt{\displaystyle \frac{\rm Ma}{\rm Re}}\left(2+\displaystyle\frac{2}{S^2}+\displaystyle\frac{1.058}{S}\sqrt{\frac{T_{\rm d}}{T_{\rm g}}}-\displaystyle\frac{1}{S^4}\right)}{1+1.86\displaystyle\sqrt{\frac{\rm Ma}{\rm Re}} }, 
\end{equation}    
where
\begin{equation}
    S=\rm{Ma}\displaystyle\sqrt{\frac{\gamma}{2}},
\end{equation}
$T_{\rm d}/T_{\rm g}$ is the ratio of the dust and gas temperatures. Intermediate values for the drag coefficient (when $1<\rm{Ma}<1.75$) can be obtained using a linear interpolation:
\begin{equation}
\label{eq:HendersonTrans}
  C_{\rm D}=C_{\rm D} ({\rm Ma=1})+\displaystyle \frac{4}{3}(\mathrm{Ma}-1)\left(C_{\rm D}({\rm Ma=1.75})-C_{\rm D} ({\rm Ma=1})\right),  
\end{equation}
where $C_{\rm D} ({\rm Ma=1})$ can be found from (\ref{eq:HendersonSub}), by substituting $\rm{Ma}=1$, and $C_{\rm D} ({\rm Ma=1.75})$ can be found from (\ref{eq:HendersonSuper}), by substituting $\rm{Ma}=1.75$. Here and below, we assume $\gamma=1.4$, $T_{\rm d}/T_{\rm g}=1$.

\begin{figure}
    \centering
    \includegraphics[width=180mm]{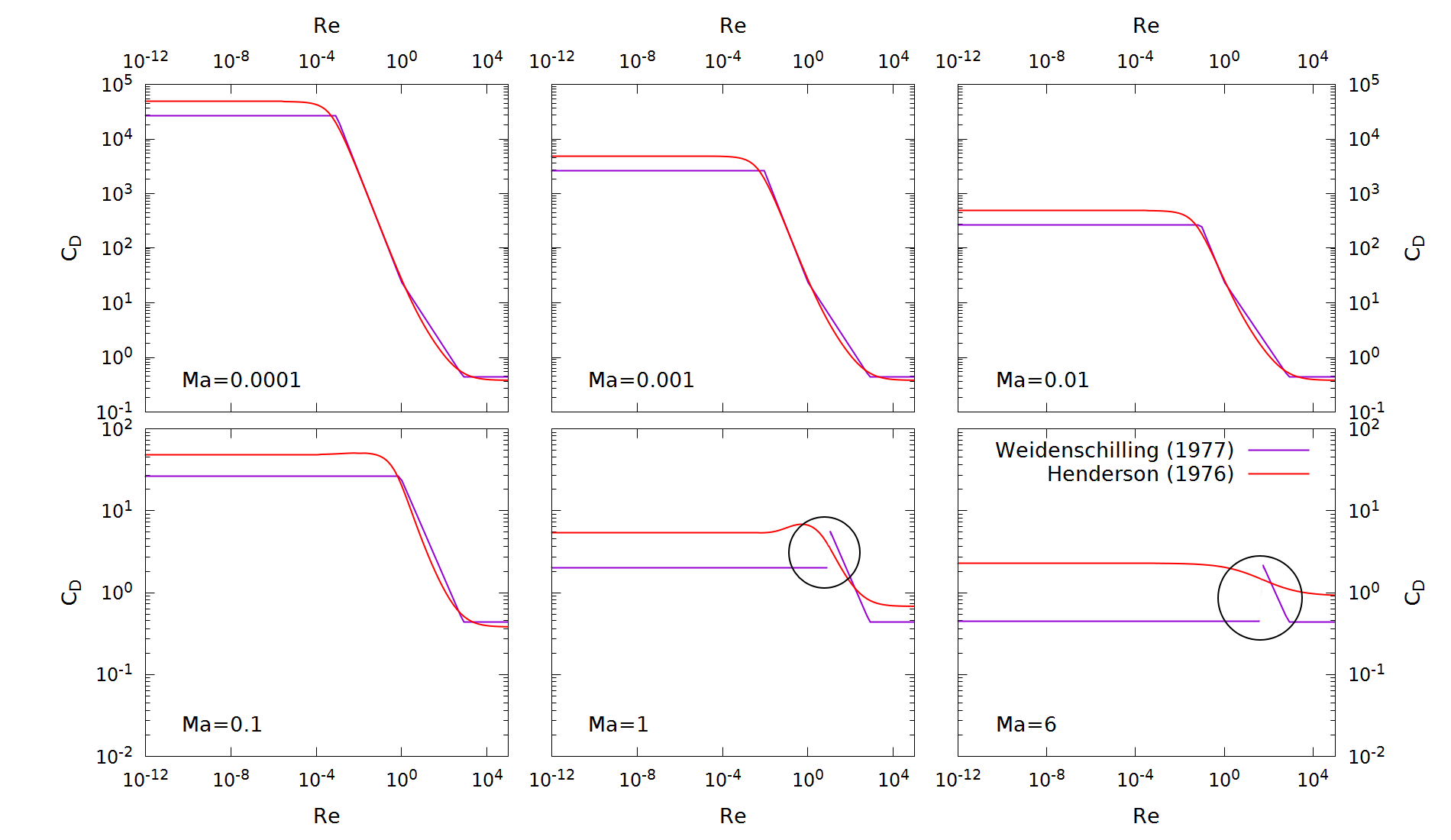}
    \caption{Henderson drag coefficient (\ref{eq:HendersonSub})-(\ref{eq:HendersonTrans}) and the standard astrophysical expression (\ref{eq:CD4}) for various $\rm{Ma}$ as a function of $\rm{Re}$.}
    \label{fig:CD_Re}
\end{figure}

\begin{figure}
    \centering
    \includegraphics[width=180mm]{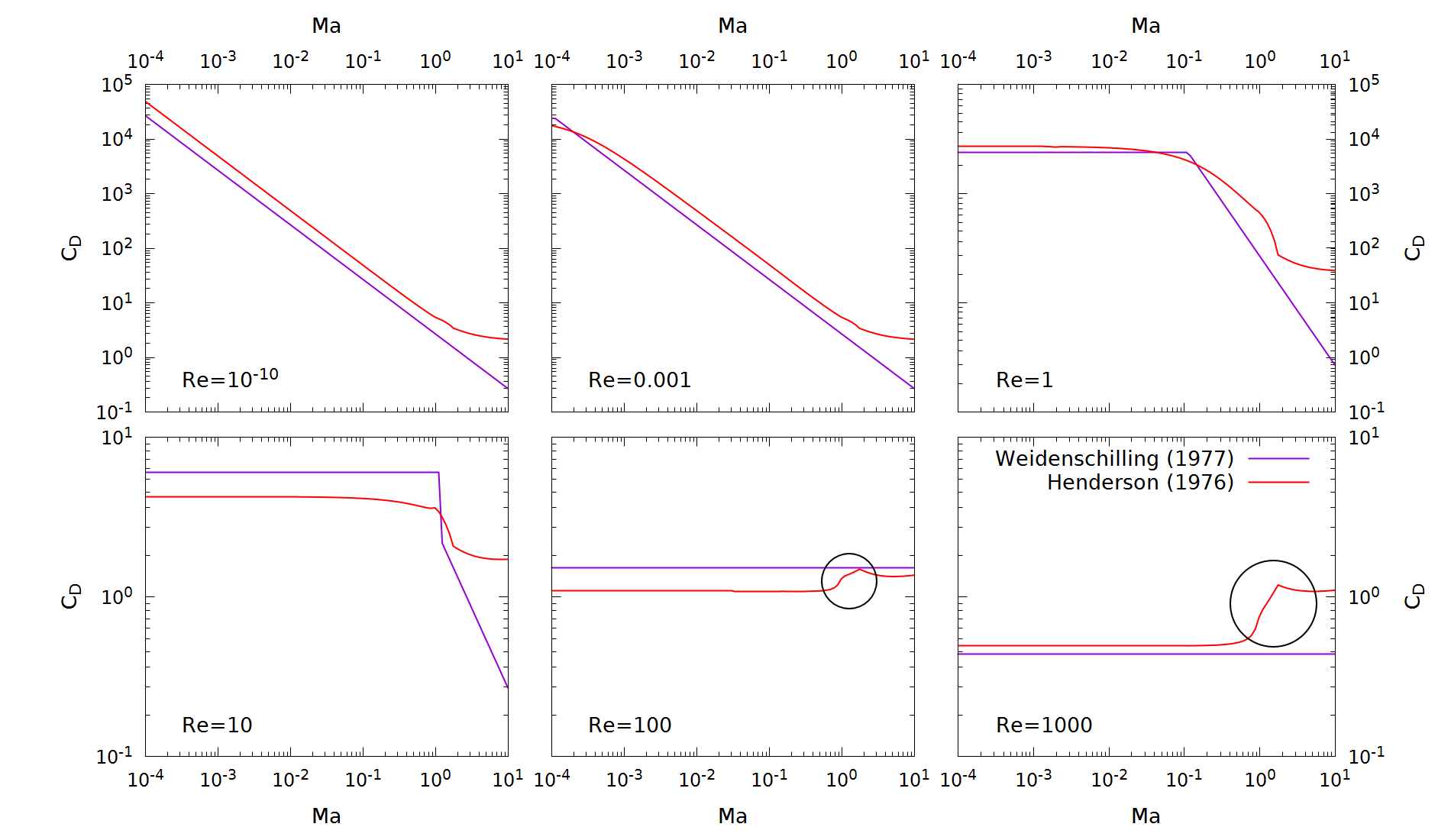}
    \caption{Henderson drag coefficient (\ref{eq:HendersonSub})--(\ref{eq:HendersonTrans}) and the standard astrophysical expression (\ref{eq:CD4}) for various  $\rm{Re}$ as a function of $\rm{Ma}$.}
    \label{fig:CD_Ma}
\end{figure}

The values of the standard astrophysical coefficient (\ref{eq:CD4}) and the Henderson coefficient (\ref{eq:HendersonSub})--(\ref{eq:HendersonTrans}) for the range of parameters of a circumstellar disk $10^{-4} \le \rm{Ma} \le 100$, $10^{-12} \le \rm{Re} \le 10^3$ are presented in Figs.\ref{fig:CD_Re} and~\ref{fig:CD_Ma}. It follows from the upper row of panels in Fig.~\ref{fig:CD_Re} that both coefficients are similar when $\rm{Ma} \le 0.01$ for any value of $\rm{Re}$. This is confirmed by the upper panels of Fig.~\ref{fig:CD_Ma}. The values of the two drag coefficients differ quantitatively when $0.1 \le \rm{Ma} \le 1$ and qualitatively when $\rm{Ma} \ge 1$. First, a discontinuity of the first type is visible in the standard astrophysical coefficient in the central and right panels of the lower row of Fig.~\ref{fig:CD_Re}. Furthermore, the lower right panel shows that the Henderson coefficient falls with growth in $\rm{Re}$, while the standard astrophysical coefficient grows. Second, the central and right panels of the lower row in Fig.~\ref{fig:CD_Ma} show that the Henderson coefficient has a feature (a bump near $\rm{Ma}=1$, well known in aeromechanics), which is absent in the simpler astrophysical approximation for the coefficient.

Thus, when simulating the dynamics of a gas–dust medium with $\rm{Ma} > 1/9$, it is recommended to use only the Henderson coefficient. However, if the treatment is restricted to $\rm{Ma} \le 1/9$, the standard astrophysical coefficient is less computationally expensive and convenient for programming.     

\section{NUMERICAL METHODS FOR SOLVING THE EQUATIONS OF MOTION OF A GAS-DUST MEDIUM WITH INTENSE INTER-PHASE INTERACTIONS}

\subsection{Review of Earlier Studies}
\label{sec:previos work}

We define the time scale for the velocity of a spherically
symmetrical particle to relax to the velocity of
the ambient gas as follows:
\begin{equation}
\label{eq:tstop}
t_{\rm stop}=\frac{m_{\rm p}\|v-u\|}{\|F_{\rm D}\|}=\frac{\frac{4}{3} \pi a^3 \rho_{\rm s} \|v-u\|}{\frac{1}{2} C_{\rm D} \pi a^2 \rho_{\rm g} \|v-u\|^2}=\frac{8}{3}\frac{a\rho_{\rm s}}{C_{\rm D} \rho_{\rm g} \|v-u\|},
\end{equation}
where $m_{\rm p}$ is the mass of the particle and $\rho_{\rm s}$ is the material
density of the grains. 

For dust grains with sizes of about one micron in an
orbit with a radius of 100 au, the estimated relaxation,
or stopping, time $t_{\rm stop}$ in the disk is of order 100 s (see,
for example, \cite{StoyanovskayaDust,LaibePrice2011Test}), while the disk dynamics require
simulations over $10^4$~yrs $\approx 10^{11}$~s or more. This means
that the equations of motion of the gas and dust have
stiff relaxation terms associated with intense interphase
interactions. This makes simulations the
dynamics of gas–dust disks a challenging task in modern
computational astrophysics (see, for example, \cite{HaworthEtAl2016}). 

Problems with stiff relaxation terms have been
actively studied in applied mathematics \cite{StiffRelaxationTerm}. In addition
to the dynamics of multiphase media, such problems
also arise in plasma physics, models for elasticity
with memory, problems in kinetic theory, and so forth.
When explicit methods are used to obtain stable
numerical solutions, the time step must satisfy the
condition $\tau \leq t_{\rm stop}$, which is unacceptably expensive
for simulations of multi-physics systems of equations in the two-dimensional and three-dimensional cases,
even on modern supercomputers.

In particular, when modeling the dynamics of a
gaseous, self-gravitating disk (e.g., \cite{Vorobyov2010,StoyanovskayaACBurst}), the time
step indicated by the Courant condition is several
Earth days. When necessary, this step can be made up
to a factor of 100 shorter, but the time for the solutions
to be obtained then increases by a factor of 100–1000,
making it infeasible to carry out series of computational
experiments. 

There is no restriction on the time step from the
stability point of view when using implicit methods,
but restrictions can arise due to the required accuracy \cite{Pember1999}. For example, in an early study, Jim and Livermore \cite{NumericsStiff} showed that, in general, in the presence of
stiff relaxation terms, even high-accuracy grid methods
can reproduce asymptotic behavior of the solutions
only if a very small spatial step is used. Examples
of the manifestation of such effects in simulations of
the dynamics of circumstellar disks can be found in \cite{StoyanovskayaDust,VorobyovEtAl2017,LaibePriceAstroDrag}.  
In particular, the diffusion overdissipation of
solutions obtained in simulations of the dynamics of a
gas–dust medium using smoothed particle hydrodynamics
(SPH) with computations of the interphase
interactions using a classical approach \cite{MonaghanKocharyan1995} were studied
in \cite{LaibePrice2014OneFluidDust,LaibePriceAstroDrag}. An empirical criterion was derived,
according to which the overdissipation of the solutions
is at an acceptable level if the smoothing radius $h<c_{\rm s} t_{\rm stop}$ is used. According to this criterion, simulations
of the dynamics of micron-sized particles in a circumstellar disk requires a spatial resolution $h \approx 10$~km, whereas the radius of the disk is of order $10^{10}$~km.
It is clear that such a spatial resolution is
beyond the capabilities of modern computational
technology.


In a number of cases, a search for a numerical solution
of a stiff problem can be substituted with the solution
to a simpler problem obtained from the original
one via expansion in a small parameter (this general
idea is presented in \cite{StiffRelaxationTerm}, and its application to simulations
of a gas–dust medium in a circumstellar disk in \cite{JohansenKlahr2005,Akimkin2015SFTA,Akimkin2017SFTA}, for example). The simplified problem
becomes non-stiff, but this approach is justified only if
the relaxation time is much smaller than the time scale
for the motion of the carrier phase, and it cannot be
used to model the transition regime when the relaxation
time is close to the dynamical time scale. Moreover,
this approach may involve a change in the type of
equations considered.

Therefore, an alternative to adopting a simplified
problem is the development of implicit and semiimplicit
numerical methods that are able to precisely
reproduce the equilibrium values, even with crude
spatial resolution. Attempts are currently being made
to find a general approach to the construction of highaccuracy
methods for problems with relaxation terms
(see, e.g., \cite{Albi2019}). However, in most studies of the
numerical solution of such systems, the methods are
constructed taking into account the specifics of the
problem to be solved and specific forms for the “stiff
relaxation term”. In particular, when simulating the
dynamics of a gas–dust medium, the system has not
one, but two stiff relaxation terms—in the equations of
motion for the gas and dust \cite{BateDust2014,LaibePrice2014OneFluidDust,Ishiki2017,YangJohansen2016}). On average
over the disk, $\varepsilon=\displaystyle \frac{\rho_{\rm d}}{\rho_{\rm g}}$ is equal to 0.01, to order of magnitude,
so that the influence of the dust on the gas
dynamics can often be neglected. On the other hand,
the results of simulations indicate that the dust particles
may be concentrated in specific regions of the disk
(e.g., in spiral arms \cite{RiceEtAl2004}, in the inner part of the disk \cite{VorobyovEtAl2017}, in self-gravitating gaseous clumps \cite{ChaNayakshinDust2011}), increasing
the density ratios in these locations to values $\varepsilon\approx 1$ or
more. 

Several groups of researchers are currently working
on creating numerical models for the dynamics of a
gas–dust medium with a high concentration of the
dispersed phase and intense interphase interactions,
based on grid methods and SPH.

Benitez-Llambay et al. \cite{BenitezLlambay2019} presented a non-iterative
implicit grid method and set of test problems for a
medium of gas and polydisperse dust in a circumstellar
disk. In their model, each fraction of the polydisperse
dust exchanges momentum with gas, but there is no
energy exchange between the gas and dust. They
showed that their proposed method is stable, has low dissipative properties, and preserves the asymptotic
behaviour of the solution for any time step.

Sadin \cite{Sadin2016} and Sadin and Odoev \cite{SadinOdoev2017} proposed a
finite-difference scheme with customizable dissipative
properties aimed at simulating the dynamics of a
medium of gas and monodisperse dust.

A non-iterative, implicit method based on SPH for
monodispersive and polydisperse dust is presented in \cite{LaibePriceOneFluidSPH,LaibePriceAstroDrag,Multigrain}. This model for the dynamics of the gas–
dust medium is based on a one-fluid approach (the
equations for the properties of the gas–dust medium
as a whole are solved, solving the diffusion equation
for the concentration of the solid phase). In this
approach, each SPH particle carries the features of
both the gas and the dust. This approach is fast, and
free of the numerical dissipation that arises in the
application of Lagrangian methods in a two-fluid
approach. On the other hand, the conservation of
mass for the dispersed phase is not guaranteed in a
one-fluid approach. In addition, there is a stricter
bound from above on the size of the dispersed phase
particles, compared to the two-fluid approach.

Lóren-Aguilar and Bate \cite{BateDust2014,BateDust2015} developed a twofluid
approach based on SPH for a medium of gas and
monodispersive dust. They found that the numerical
overdissipation in the gas–dust medium was smaller in
their solutions than in the classical SPH approach \cite{MonaghanKocharyan1995}. However, this dissipation was higher than the
diffusion dissipation in the gas due to artificial
viscosity.

Stoyanovskaya et al. \cite{IDIC2018} proposed a new method
for solving the motion equations of gas and monodisperse
dust with intense interphase interactions
based on SPH. Their proposed method for computing
the interphase interactions is non-iterative and has low
dissipative properties. These features of the method
are determined by the implicit linear approximation
for the drag force used, together with the local conservation
of momentum in the computations of the drag
between the gas and dust.

All these developments were made for the most
common regime for the interactions between particles
and gas in a circumstellar disk—the Epstein regime. In
this regime, there are intense interphase interactions,
and drag term depends linearly on the relative velocity
between the gas and particles. However, the simulation
results show that grown solids interact with the gas
in a transition and Newton regimes, when the interphase
interactions are moderate and the drag depends
non-linearly on the relative velocity between the gas
and particles. Accordingly, in our current study, we
propose a scheme for the time discretization of the
equations of motion of the gas and dust that encompasses
both cases and can be used with any method for
computing the spatial derivatives if the velocities of the
gas and dust are determined at the same points in
space. This method for the simulation of the dynamics
of a medium of gas and dust with drag that depends linearly on the relative velocity between the gas and
dust is a special case of the approach of \cite{BenitezLlambay2019}, which is
well justified in the linear case. Thus, our study can be
considered as a basis for the approach to the simulation
of polydisperse dust \cite{BenitezLlambay2019} with a transition from
linear to non-linear drag.

\subsection{Fast and Asymptotic Preserving Scheme
for the Time Discretization of the Equations
of Motion of the Gas and Dust}
\label{sec:scheme}
To approximate the interphase interactions in the
equations of motion in (\ref{eq:gas})-(\ref{eq:dust}), we linearized the drag
term on the relative velocity between the gas and particles,
with the stopping time $t_{\rm stop}(\rho_{\rm g},u-v)$ computed
using known quantities from the previous step and the
relative velocity of the gas and particles computed
from the subsequent step. A simple description of this
approach has the form:
\begin{equation}
\label{eq:SchemeGas}
 \displaystyle\frac{v^{n+1}-v^n}{\tau}+(v^n \cdot \nabla) v^n =-\displaystyle\frac{\nabla p^n}{\rho_{\rm g}^n}+ g^n - \frac{\rho^n_{\rm d}}{\rho^n_{\rm g}} \frac{v^{n+1}-u^{n+1}}{t^n_{\rm stop}}  +\frac{f^n_{\rm g}}{\rho_{\rm g}^n},
\end{equation}
\begin{equation}
\label{eq:SchemeDust}
\displaystyle\frac{u^{n+1}-u^n}{\tau}+(u^n \cdot \nabla) u^n = g^n + \frac{v^{n+1}-u^{n+1}}{t^n_{\rm stop}} + \frac{f^n_{\rm d}}{\rho_{\rm d}^n}.
\end{equation}

We solved the entire system (\ref{eq:gas})-(\ref{eq:dust}) using a two-stage
scheme based on the operator splitting method
with respect to physical processes (as in our earlier
studies \cite{StoyanovskayaDust,Development2018}). We solved the continuity equation
and the equation of motion using the finite-difference
method described in detail for a pure gas in \cite{StoneNorman1992}. In the
first stage of the scheme, the advective terms and
transport of mass and momentum are computed:
\begin{equation}
\label{adv:gas}
\displaystyle\frac{\partial \rho_{\rm g}}{\partial t}+\nabla (\rho_{\rm g} v)=0,\ \ \ 
\rho_{\rm g} \left[\displaystyle\frac{\partial v}{\partial t}+(v \cdot \nabla) v \right]=0,
\end{equation}
\begin{equation}
\label{adv:dust}
\displaystyle\frac{\partial \rho_{\rm d}}{\partial t}+\nabla (\rho_{\rm d} u)=0,\ \ \ 
\rho_{\rm d} \left[\displaystyle\frac{\partial u}{\partial t}+(u \cdot \nabla) u \right]=0.
\end{equation}
We computed the spatial derivatives in each stage
using the piecewise-parabolic method of \cite{ColellaWoodward1984}. In this
stage, we determined the quantities $\rho_{\rm g}^{n+1/2}$, $\rho_{\rm d}^{n+1/2}$, $v^{n+1/2}$, $u^{n+1/2}$.

In the next stage, we computed the influence of the
forces of pressure, drag, gravitation, and other effects
using the updated densities and velocities of the gas
from the first stage:
\begin{equation}
\label{eq:SourceGas}
 \displaystyle\frac{v^{n+1}-v^{n+1/2}}{\tau} =-\displaystyle\frac{\nabla p^{n+1/2}}{\rho_{\rm g}^{n+1/2}}+ g^{n+1/2} - \frac{\rho^{n+1/2}_{\rm d}}{\rho^{n+1/2}_{\rm g}} \frac{v^{n+1}-u^{n+1}}{t^{n+1/2}_{\rm stop}}  +\frac{f^{n+1/2}_{\rm g}}{\rho_{\rm g}^{n+1/2}},
\end{equation}
\begin{equation}
\label{eq:SourceDust}
\displaystyle\frac{u^{n+1}-u^{n+1/2}}{\tau} = g^{n+1/2} + \frac{v^{n+1}-u^{n+1}}{t^{n+1/2}_{\rm stop}} + \frac{f^{n+1/2}_{\rm d}}{\rho_{\rm d}^{n+1/2}}.
\end{equation}

Thanks to the fact that $v^{n+1}$ and $u^{n+1}$ can be
expressed explicitly using (\ref{eq:SchemeGas}), (\ref{eq:SchemeDust}), the computational
costs required for the semi-implicit approximation
of friction are similar to those for an explicit
approximation. 

\subsection{Testing of the Scheme}
\label{sec:test}

In Section (\ref{sec:DustyShock}), we will consider the degree of
numerical overdissipation of the scheme (\ref{eq:SchemeGas}), (\ref{eq:SchemeDust}) during the computation of intense interphase interaction
using a time step $\tau$, determined by the Courant
condition, and not $t_{\rm stop}$. In Section \ref{sec:toymodel} we will compare
the accuracy of solutions obtained using the firstorder
scheme (\ref{eq:SchemeGas}), (\ref{eq:SchemeDust}) and a fourth-order scheme to
simulate the motion of single dust grain in a gaseous
disk whose angular velocity is slightly sub-Keplerian.

\subsubsection{Test 1. Strong shocks in a gas–dust medium.} 
\label{sec:DustyShock}


In the one-dimensional equations (\ref{eq:gas}) and (\ref{eq:dust}), we set $g=0$, $S_{\rm g}=S_{\rm d}=0$, $f_{\rm g}=f_{\rm d}=0$, $f_{\rm D}=\displaystyle \frac{ \rho_{\rm d}(v-u)}{t_{\rm stop}}$ and
wrote the energy equation:

\begin{equation}
\label{eq:ShockWaveEnergyGas}
\rho_{\rm g} \left(\frac{\partial e}{\partial t} +v\frac{\partial e}{\partial x}\right)=-p\frac{\partial v}{\partial x},
\end{equation}
where $e$ is the internal energy of the gas, which is
related to the pressure as
\begin{equation}
\label{eq:EOS}
p=\rho_{\rm g} e (\gamma-1),
\end{equation}
where $\gamma$ is the adiabatic index. We made this system
dimensionless using the length $l_*$, mass $m_*$, and time $t_*$units, so that the derivative dimensional quantities
have the form $\rho_*=\displaystyle\frac{m_*}{l_*^3}$, $v_*=\displaystyle\frac{l_*}{t_*}$, $p_*=\displaystyle\frac{m_*}{t_*^2 l_*}$, $e_*=\displaystyle \frac{l_*^2}{t_*^2}$. We
solved the system in the dimensionless variables.

For the resulting system (\ref{eq:gas}), (\ref{eq:dust}), (\ref{eq:ShockWaveEnergyGas}), (\ref{eq:EOS}), we
specified a no-flow condition at the boundaries of an
interval. We specified the velocities to be zero at the
initial time, and specified the jumps in the gas pressure
and in the densities of the gas and dust to be

 $$\left[\displaystyle\frac{\rho_{\rm{g}}}{\rho_*}, \frac{p}{p_*}, \frac{e}{e_*}, \frac{\rho_{\rm{d}}}{\rho_*}\right]_{\it l}=[1, 1, 2.5, 1],$$
 $$\left[\displaystyle\frac{\rho_{\rm{g}}}{\rho_*}, \frac{p}{p_*}, \frac{e}{e_*}, \frac{\rho_{\rm{d}}}{\rho_*}\right]_{\it r}=[0.125, 0.1, 2, 0.125],$$

 $$\gamma_l = \gamma_r = 7/5.$$


We specified $\lambda=\displaystyle 5 \cdot 10^{-6} l_* \frac{\rho_*}{\rho_{\rm g}}$, $\rho_{\rm s}=2.3 \rho_*$ and considered
two particle sizes: small grains with $a=5 \cdot 10^{-6} l_*$ and large grains with $a=10^{-2} l_*$. 


\begin{figure}
    \centering
    \includegraphics[width=190mm]{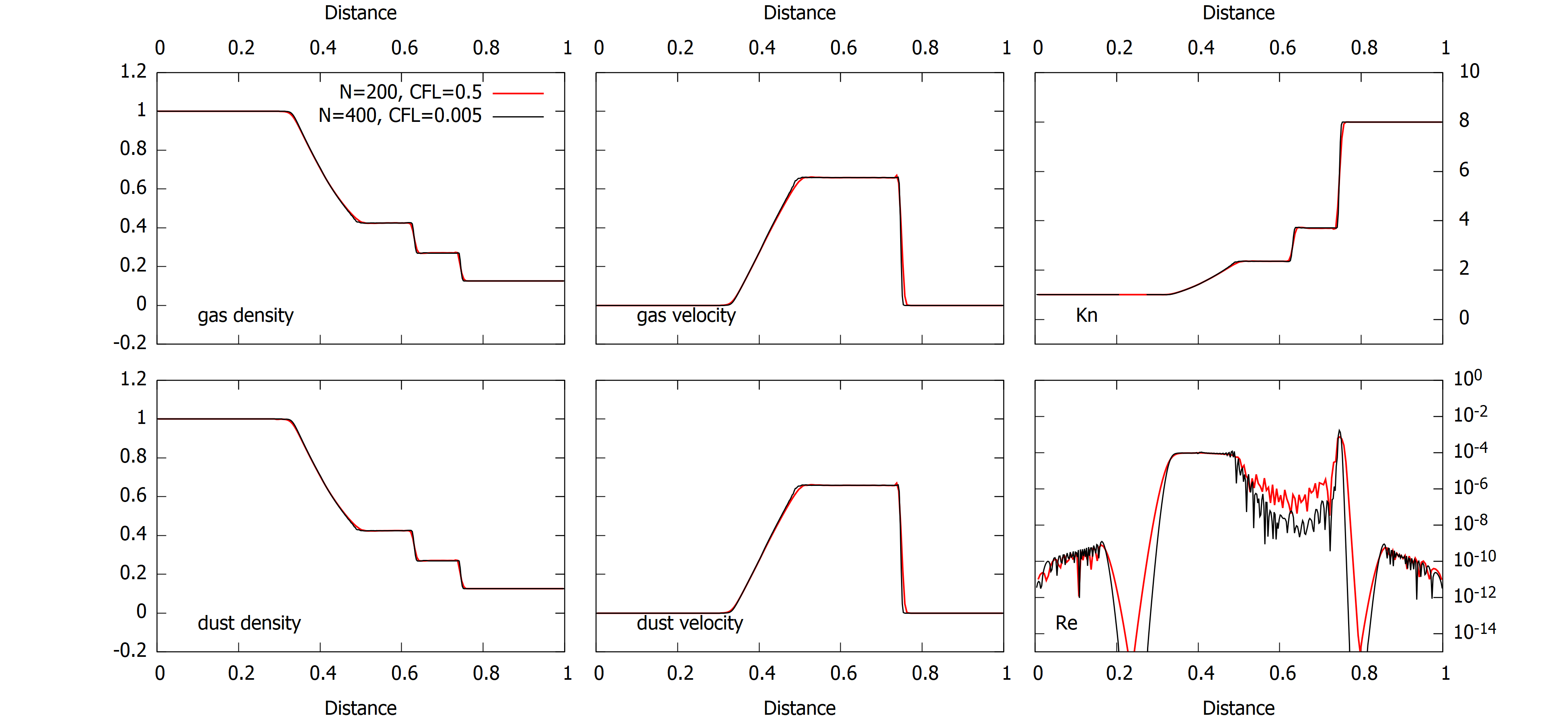}
    \caption{Solution for a Sod shock tube for the scheme (\ref{eq:SchemeGas}), (\ref{eq:SchemeDust}), which preserves the asymptotic behavior, for $\rm{Kn} \geq 1$ (small dust
with $a=5 \cdot 10^{-6} l_*$, the Epstein regime, Henderson drag coefficient (\ref{eq:HendersonSub})--(\ref{eq:HendersonTrans})).}
    \label{fig:DustyShockSmall}
\end{figure}

\begin{figure}
    \centering
    \includegraphics[width=190mm]{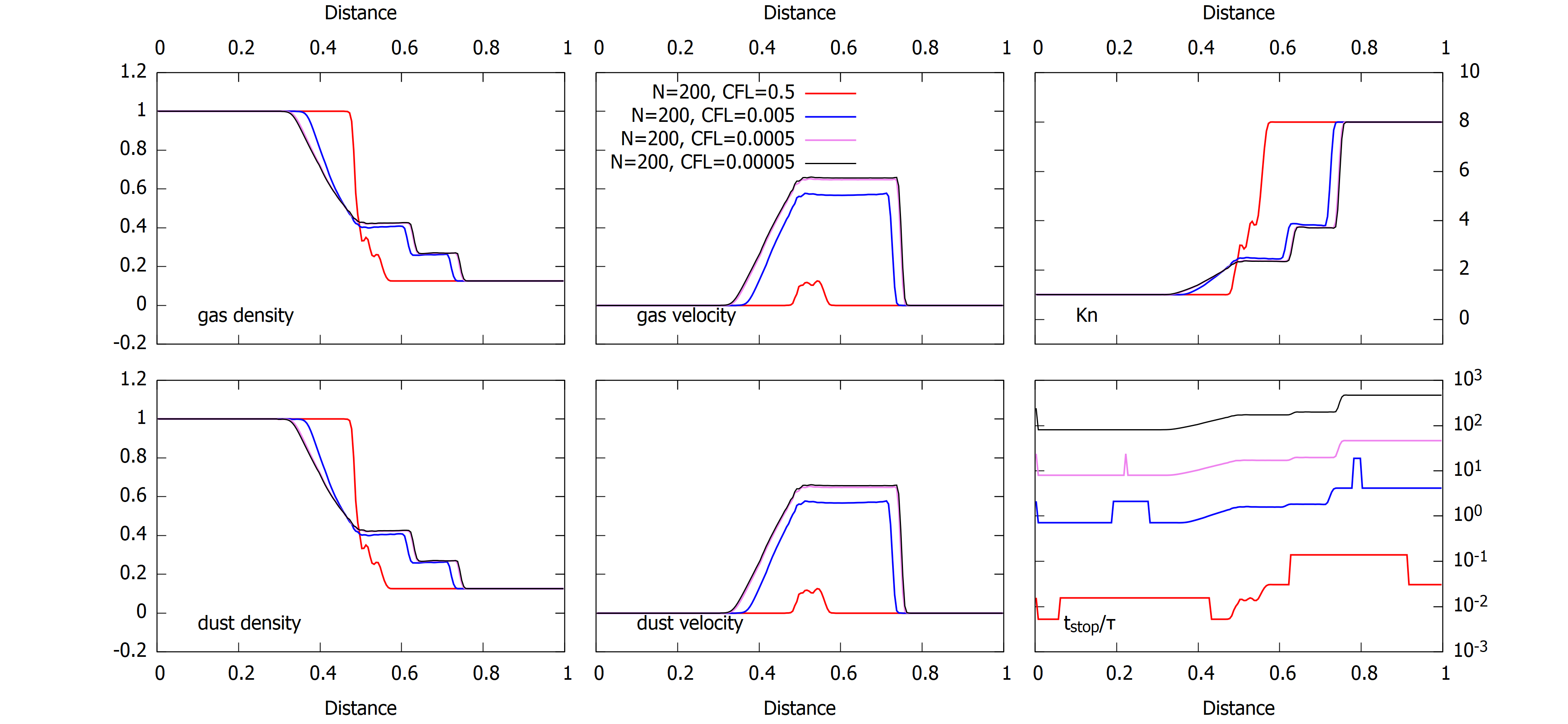}
    \caption{Solution for a Sod shock tube for the scheme (\ref{eq:SchemeGasNA}), (\ref{eq:SchemeDustNA}), which does not preserve the asymptotic behavior, with various
time steps for $\rm{Kn} \geq 1$ (small dust with $a=5 \cdot 10^{-6} l_*$, Epstein regime, Henderson drag coefficient (\ref{eq:HendersonSub})--(\ref{eq:HendersonTrans})).}
    \label{fig:OLDDustyShockSmall}
\end{figure}

It was shown in \cite{StoyanovskayaDust} that the semi-implicit, first-order operator splitting scheme (\ref{eq:SchemeGas}), (\ref{eq:SchemeDust}) preserves the asymptotic solutions in computations of intense interphase interactions with a constant drag coefficient. This same conclusion follows from Fig..~\ref{fig:DustyShockSmall} for the case when $C_{\rm D}$ varies in space and time and $t_{\rm stop}=t_{\rm stop}(\|u-v\|)$. 

Figure~\ref{fig:DustyShockSmall} presents simulations with the Henderson drag coefficient (\ref{eq:HendersonSub})--(\ref{eq:HendersonTrans}) for a grain size $a=5 \cdot 10^{-6}l_*$. We adopted the results for our simulations with detailed spatial resolution with $N=400$ cells per segment and a time step for which the Courant parameter is $\mathrm{CFL}=0.005$ as a reference. It is clear that the computations of the gas and dust velocities and densities with $N=200, \mathrm{CFL}=0.5$ differ only slightly from the reference results. In addition to computations with $\rho_{\rm d} / \rho_{\rm g}=1$, we also carried out simulations for the same problem with the dust content relative to the gas content enhanced by a factor of 10 and 1000. We confirmed that the level of numerical overdissipation of the solution remained insignificant for these cases in computations using the asymptotic-preserving scheme.

An example of a semi-implicit drag term approximation that does not preserve the asymptotic behavior is
\begin{equation}
\label{eq:SchemeGasNA}
 \displaystyle\frac{v^{n+1}-v^n}{\tau}+(v^n \cdot \nabla) v^n =-\displaystyle\frac{\nabla p^n}{\rho_{\rm g}^n}+ g^n - \frac{\rho^n_{\rm d}}{\rho^n_{\rm g}} \frac{v^{n+1}-u^{n}}{t^n_{\rm stop}}  +\frac{f^n_{\rm g}}{\rho_{\rm g}^n},
\end{equation}
\begin{equation}
\label{eq:SchemeDustNA}
\displaystyle\frac{u^{n+1}-u^n}{\tau}+(u^n \cdot \nabla) u^n = g^n + \frac{v^{n+1}-u^{n+1}}{t^n_{\rm stop}} + \frac{f^n_{\rm d}}{\rho_{\rm d}^n}.
\end{equation}

The solution of this problem obtained using the scheme (\ref{eq:SchemeGasNA}), (\ref{eq:SchemeDustNA}) is presented in Fig.~\ref{fig:OLDDustyShockSmall}. This figure
shows that, when $\mathrm{CFL}$ is varied from 0.5 to 0.00005,
the computed wave-propagation speeds vary significantly,
with the accuracy of the resulting solution
being satisfactory only when $\tau<0.1 t_{\rm stop}$.

\begin{figure}
    \centering    
    \includegraphics[width=120mm]{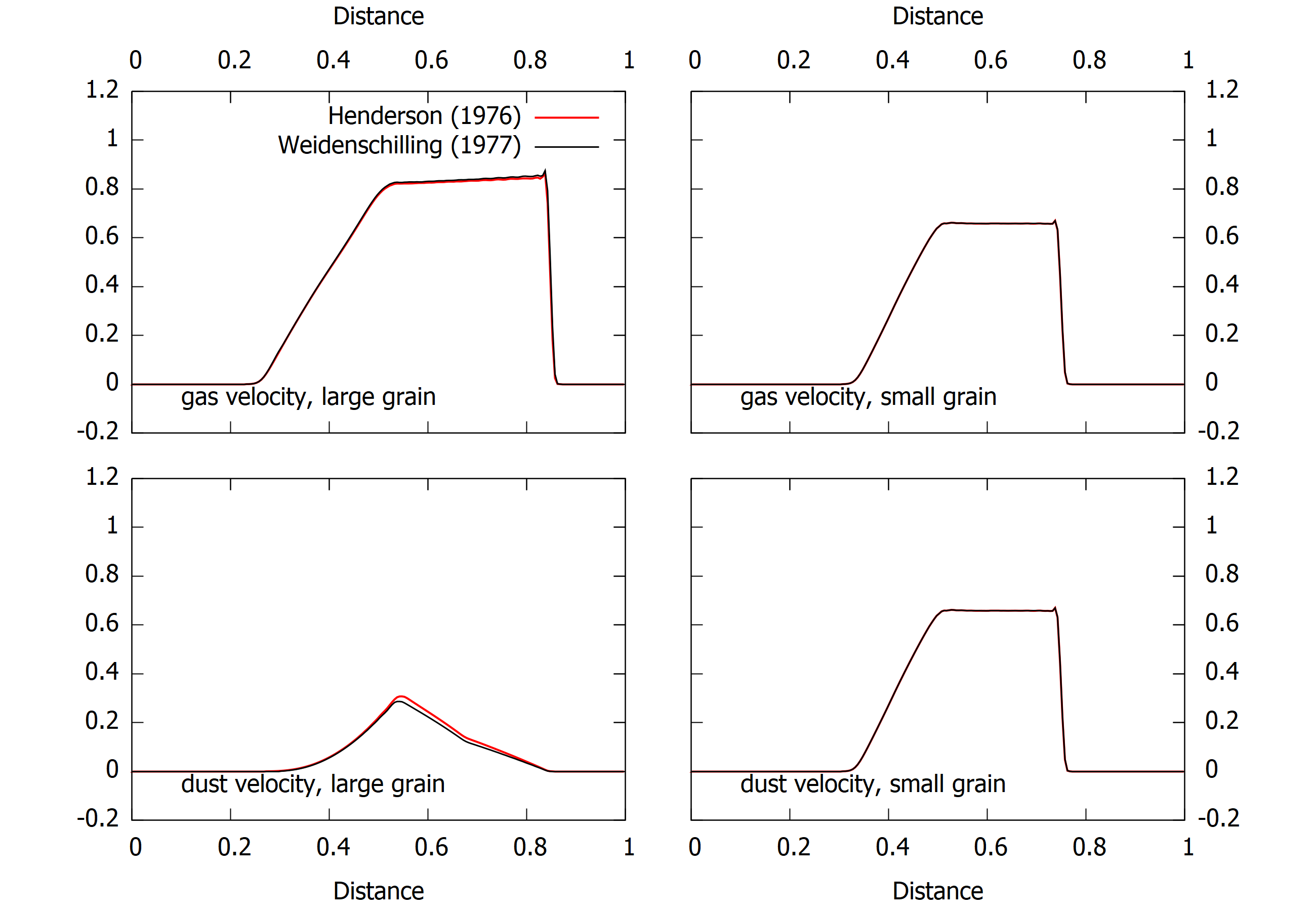}
    \caption{Solution for a Sod shock tube for large bodies (left panels) and small grains (right panels) with the Henderson (\ref{eq:HendersonSub})--(\ref{eq:HendersonTrans}) and standard astrophysical (\ref{eq:CD4}) drag coefficients.}
    \label{fig:DustyShock2CD}
\end{figure}

It follows from the upper panel of Fig.~\ref{fig:DustyShockSmall} that the
Knudsen number for the small grains exceeds unity.
This means that, according to the regime classification (\ref{eq:CD4}), the interaction of the gas and dust occurs in
the Epstein free-molecular-flow regime. The computation
of $t_{\rm stop}^n$ was carried out using the general formula (\ref{eq:tstop}) for linear and non-linear drag term. Note
that the influence of the numerical resolution on the
results is visible only in the lower right panel of Fig.~\ref{fig:DustyShockSmall}, which presents the quantity $\rm{Re}$, which depends linearly
on $\|u-v\|$. It is known, however (see, e.g., \cite{LaibePrice2011Test}), that the relative velocity of the gas and dust can be
ignored in the case $t_{\rm stop} \max(u,v) \ll l_*$, and the solution for a gas–dust medium can be obtained from the
solution for a gaseous medium by replacing $c_{\rm s}$ by
\begin{equation}
\label{eq:dustysound}
c^*_s=c_s\left(1+\displaystyle\frac{\rho_{\rm d}}{\rho_{\rm g}}\right)^{-1/2}.
\end{equation}
This means that the solution of the stationary problem
does not depend on the drag coefficient and $t_{\rm stop}$, and
is determined by the ratio of the dust and gas densities.
Therefore, the appreciable deviations of $\rm{Re}$ for different
spatial resolutions presented in the lower right
panel of Fig.~\ref{fig:DustyShockSmall}, do not lead to substantial differences in
the quantities depicted in the other panels. This conclusion
is supported by the right panels of Fig.~\ref{fig:DustyShock2CD}, which show the velocities of the gas and dust obtained
for small grains with $N=200$ and $\mathrm{CFL}=0.5$ and for
various drag coefficients.

\begin{figure}
    \centering
    \includegraphics[width=170mm]{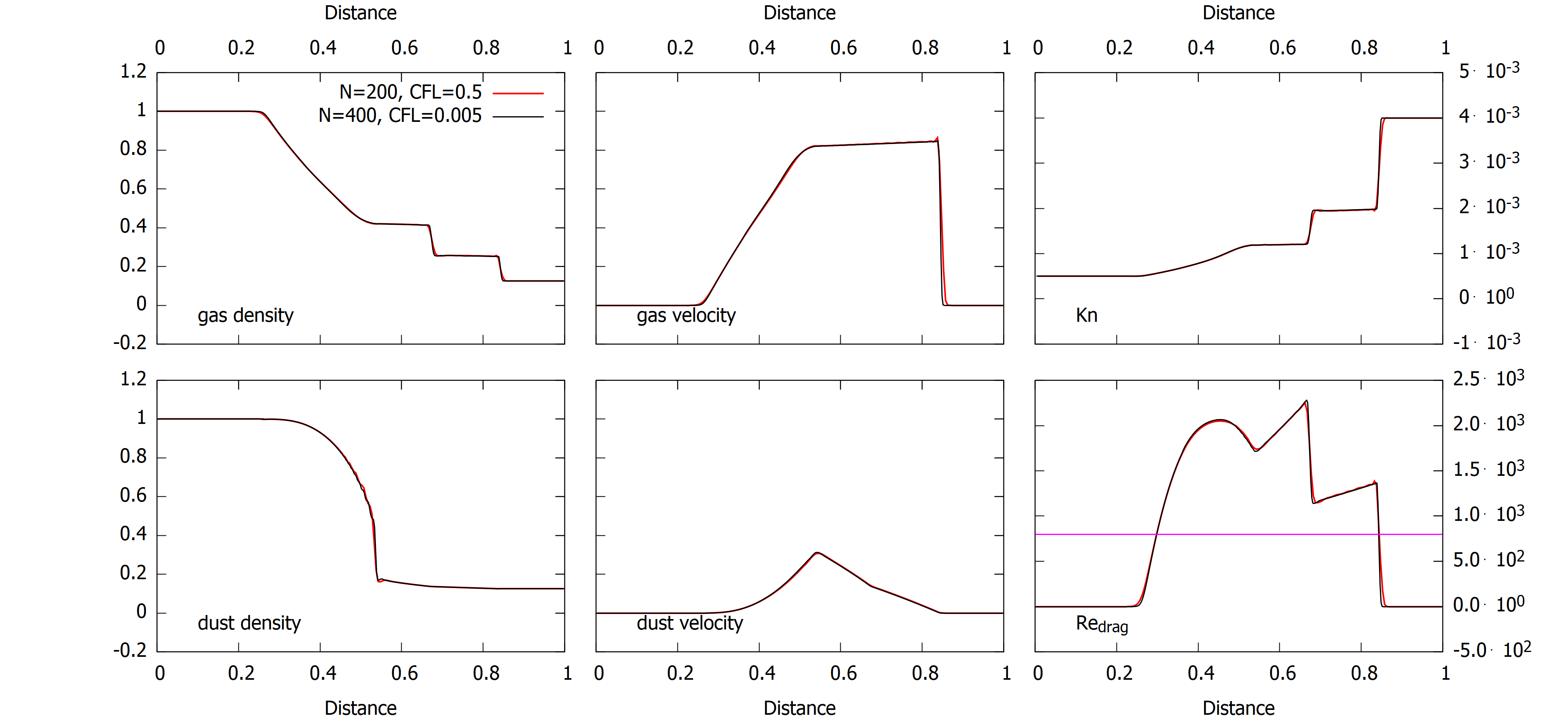}
    \caption{Solution for a Sod shock tube for the scheme (\ref{eq:SchemeGas}), (\ref{eq:SchemeDust}), which preserves the asymptotic behavior, with various steps in
space and time, for $\rm{Kn} \ll 1$ (large bodies, the Stokes, transition, and Newton regimes, non-linear drag), Henderson drag coefficient (\ref{eq:HendersonSub})--(\ref{eq:HendersonTrans})).}
    \label{fig:DustyShockLarge}
\end{figure}

The level of numerical overdissipation of the solution
obtained using the scheme (\ref{eq:SchemeGas}), (\ref{eq:SchemeDust}) for the case
of moderate interphase interactions can be seen in Fig.~\ref{fig:DustyShockLarge}, which shows the same computational results as
in Fig.~\ref{fig:DustyShockSmall} for an enhanced grain size of $a=10^{-2}l_*$. It follows
from the right panels of Fig.~\ref{fig:DustyShockLarge} that, for the specified
grain size, the gas flows around the particles like a
continuous medium ($\rm{Kn} \ll 1$) with the Stokes regime,
transition regime, and Newton regime being realized.
The simulation results for $N=200, \mathrm{CFL}=0.5$ differ
only insignificantly from the reference solution. In
contrast to the regime with intense interphase interactions,
the velocity and density distribution of the dust
differ appreciably from those for the gas. In the left panels of Fig.~\ref{fig:DustyShock2CD}, which show the velocities of the gas
and dust, we can see a more clearly defined dependence
of the solution on the drag coefficient for larger
than for smaller grains.

Overall, the results presented show that the
scheme (\ref{eq:SchemeGas}), (\ref{eq:SchemeDust}) preserves the asymptotic behavior of
the solution in simulations with both intense and
moderate interphase interactions. 

\subsubsection{Test 2. Simulation of a particle trajectory in a
stationary circumstellar disk.}
\label{sec:toymodel}

The equation of motion
of a particle in the field of a massive central body in
polar coordinates is presented, for example, in \cite{Landavshitz6}. If
gas drag acts on the particle together with gravitational
and the centrifugal force, according to (\ref{eq:DragForce}), these equations
acquire the form:
\begin{equation}
\label{eq:system}
\left\{
 \begin{array}{lcl}
    
        \displaystyle 
        \frac{dr}{dt} = v_r, \\
        \displaystyle 
        \frac{dv_r}{dt} =  \frac{v_{\varphi}^2}{r} - \frac{GM}{r^2} - \frac{3}{8}\frac{C_{\rm D}\rho_{\rm g}(u_r-v_r)||u-v||}{a\rho_s},\\
        \displaystyle 
        \frac{dv_{\varphi}}{dt} = - \frac{v_r v_{\varphi}}{r} - \frac{3}{8}\frac{C_{\rm D}\rho_{\rm g}(u_{\varphi}-v_{\varphi})||u-v||}{a\rho_{\rm s}},
    \end{array}
\right.
\end{equation}
where $r$ is the orbital radius of the particle, $(v_r,v_{\varphi})$ the
radial and azimuthal velocities of the particle, $(u_r,u_{\varphi})$ the velocity of the gas, $M$ the mass of the central body,
and $G$ the gravitational constant. We took the distributions
of the density, temperature, and velocity in the
axially symmetric gaseous disk to be given by the
model \cite{RiceEtAl2004} (a detailed description of the model is presented
in the Appendix B), and to display the following dependences on the radius: $\rho_{\rm g} \sim r^{-2.25}$, $c_s \sim r^{-0.75}$, $u_r=0$. Let us supposed that the central body has a
mass of $0.5$~$M_{\odot}$ and the disk extends from 1 au to
100 au and has a mass of $0.4$~$M_{\odot}$. Let the material density
of solid bodies be $\rho_{\rm s} = 2.2$~g~cm~$^{-3}$.

At the initial time, we specified the particle to have
an orbital radius $r_0=10$~au and a velocity $v_{\varphi 0}=0.25 V_{\rm K}(r_0)$, where $V_{\rm K}(r)=\displaystyle\sqrt{\frac{GM}{r}}$,
$v_{r0}=-10^{-4}$~au$\cdot \Omega(r=1$~au), where $\Omega(r)=\displaystyle\frac{V_{\rm K}(r)}{r}$. 
We assumed that the
size of particles drifting in the disk grows as $a=a_0\left(\displaystyle\frac{r}{r_0}\right)^{-3}$, with the particles a distance $r_0$ from the
central body having sizes $a_0=134.64$~cm (Toy Model 1) or $a_0=13.464$~cm (Toy Model 2). We simulated
the motion of a particle toward the central body
until its orbital radius reached 1 au. We used the semi-implicit
first-order scheme
\begin{equation}
  \label{eq:system}
\left\{
 \begin{array}{lcl}
        \displaystyle 
        \frac{r^{n+1}-r^n}{\tau} = v^n_r, \\
        \displaystyle 
        \frac{v^{n+1}_r-v^n_r}{\tau} =  \frac{(v^n_{\varphi})^2}{r^n} - \frac{GM}{(r^n)^2} - \frac{3}{8}\frac{C^n_{\rm D}\rho^n_{\rm g}(u^{n+1}_r-v^{n+1}_r)||u^n-v^n||}{a^n\rho_s},\\
        \displaystyle 
        \frac{v^{n+1}_{\varphi}-v^n_{\varphi}}{\tau} = - \frac{v^n_r v^n_{\varphi}}{r^n} - \frac{3}{8}\frac{C^n_{\rm D}\rho^n_{\rm g}(u^{n+1}_{\varphi}-v^{n+1}_{\varphi})||u^n-v^n||}{a^n\rho_{\rm s}},
    \end{array}  
\right.    
\end{equation}
with a time step of $\tau=0.01 \Omega^{-1}(r=1$~au). This value
of $\tau$ means that a particle that moves with Keplerian
velocity in a circular orbit of radius 1 au makes a full
orbit in 628 steps. This time step will be obtained from
the Courant condition when the dynamics of a gaseous
disk is modeled in cylindrical coordinates with a
resolution of 256 cells in azimuth and with Courant
number $\mathrm{CFL}=0.4$.

\begin{figure}
    \centering
    \includegraphics[width=150mm]{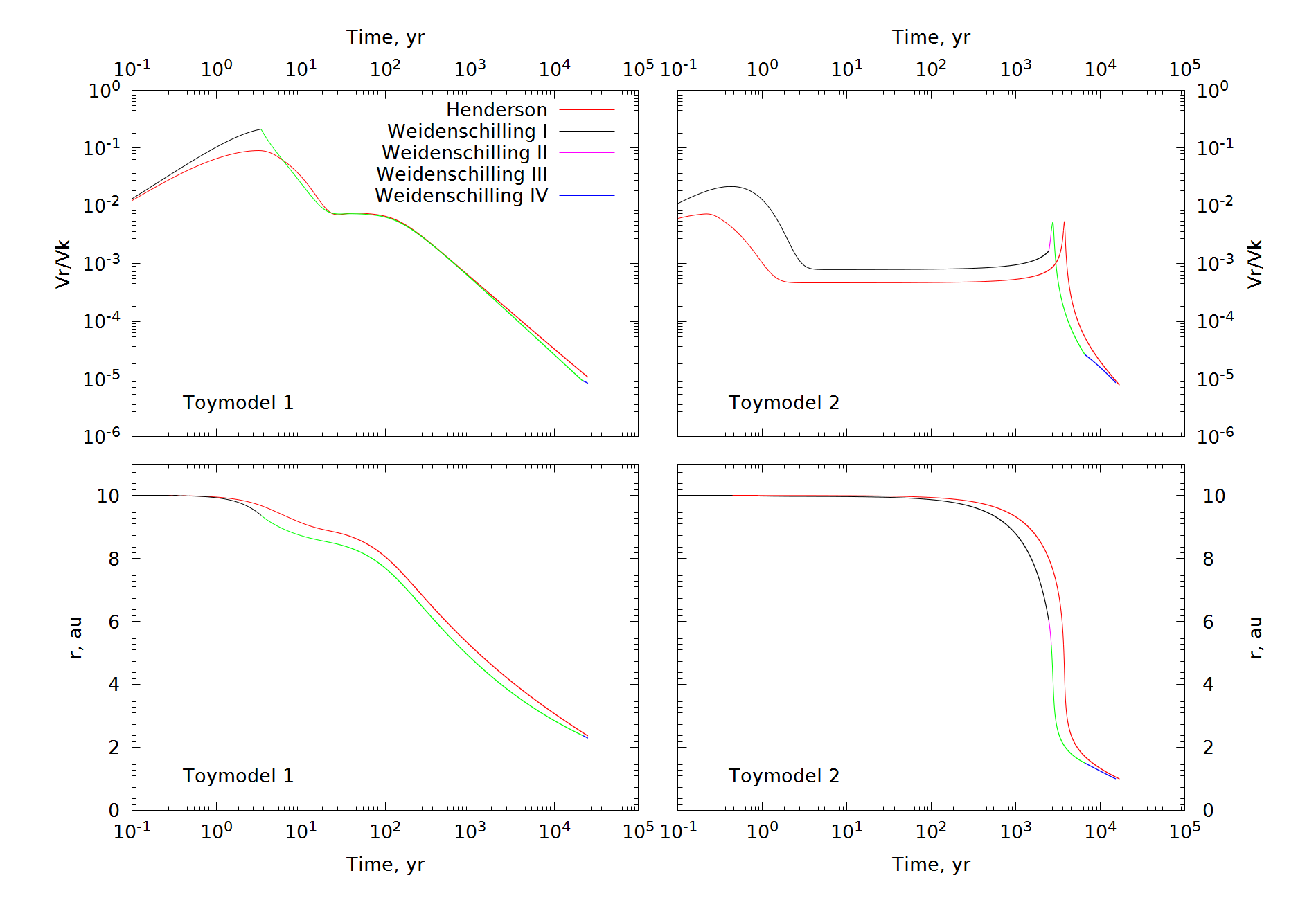}
    \includegraphics[width=150mm]{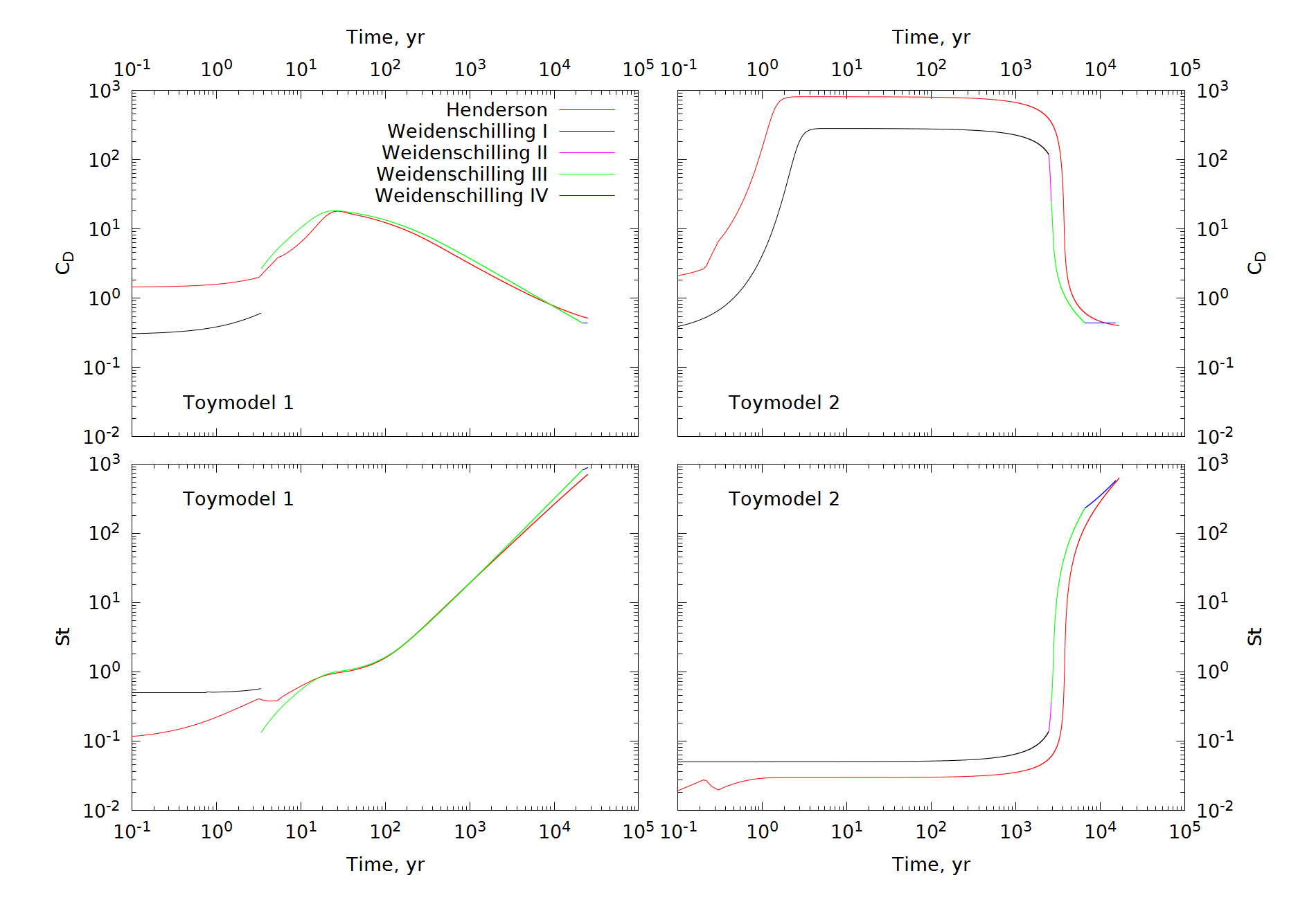}
    \caption{Simulations of the trajectory of a growing dust particle in a stationary gaseous disk with mass 0.4~$M_{\odot}$ around a central
body with mass 0.5~$M_{\odot}$ with various drag coefficients. The red curve shows the Henderson drag coefficient (\ref{eq:HendersonSub})--(\ref{eq:HendersonTrans}), he multicolored
curve the standard astrophysical coefficient (\ref{eq:CD4}) with the various regimes I–IV indicated.}
    \label{fig:1234-134}
\end{figure}

The drag coefficient $C_{\rm D}$, Stokes number $\mathrm{St}=t_{\rm stop}\Omega$, and radial component of the particle velocity $v_r$, relative to the Keplerian velocity via radius are presented in Fig.~\ref{fig:1234-134}. for Toy Models 1 and 2. The red curve shows the results obtained with the Henderson coefficient and the multicolored curve the results obtained with the standard astrophysical drag coefficient. The upper panels of Fig.~\ref{fig:1234-134} show that, in Toy Model 1, the standard astrophysical drag coefficient undergoes a discontinuity of the first type in the first ten years. The discontinuity point coincides with an extremum of the radial velocity. The radial velocity also changes from growing to decreasing in the first ten years of motion of the particle in Toy Model 1 using the continuous Henderson coefficient. The discontinuity points in Toy Models 1 and 2, which are extrema or inflection points of $C_{\rm D}$, are also extrema or inflection points of $v_r$. 

On the whole, apart from the discontinuity in $C_{\rm D}$ the simulations for Toy Model 1 with the different drag coefficients are qualitatively and quantitatively similar: the times over which the particle drifts from 10 au to 1 au are of order $3 \times 10^4$~yrs, with the Stokes number reaching $\mathrm{St}=10^3$. at a radius of 1 au. We can see that, over the time of its motion in the disk, the particle
interacts with the gas in the Epstein I, transition III,
and Newton IV regimes in (\ref{eq:CD4}), bypassing the Stokes II
regime. The lower panels of Fig.~\ref{fig:1234-134} show that, with
decreasing grain size in Toy Model 2, the drag coefficient
does not change at early times [by virtue of (\ref{eq:CD4}), it
is determined by $\rm{Ma}$ in the Epstein regime], however
the Stokes number changes by an order of magnitude
compared to Toy Model 1. As a result, the time for the
particle to drift from 10 au to 1 au is shortened, with
the particle successively passing through regimes I, II, III, and IV during its interaction with the gas. The
computational results for Toy Model 2 with the standard
astrophysical drag coefficient and with the Henderson
coefficient are qualitatively and quantitatively
similar.

As in Toy Model 1, the maximum differences in the
radial velocity when using the different drag coefficients
arise in the Epstein regime. This differs from the
shock tube problem, where the sensitivity of the solution
to changes in the drag coefficient were more
clearly expressed for large bodies in the Stokes, transition,
and Newton regimes. The sensitivity of the system (\ref{eq:system}) to a change in the coefficient of friction in the
Epstein regime is determined by the fact that the stationary
value of the radial velocity of the particle
depends on $t_{\rm stop}$ \cite{Nakagawa1986} as
\begin{equation}
\frac{v_r}{V_{\rm K}} = - {\eta \over \mathrm{St}(r) + \mathrm{St}(r)^{-1}}\;,
\label{eq:analyt}
\end{equation}
where $\eta$ is determined from $u^2_{\varphi}=(1-\eta) V^2_{\rm K}$ and satisfies $0 < \eta \ll 1$. The upper panels of Fig.~\ref{fig:CD_Re} show that, in
the Epstein regime, the Henderson coefficient differs
from the standard astrophysical coefficient by approximately
a factor of three, which, by virtue of (\ref{eq:analyt}), leads
to a substantial difference in the radial velocity of the
particle.

Two quantities are present in the second equation
of (\ref{eq:system}) whose difference is many orders of magnitude
less than their magnitudes ($\displaystyle{v_{\varphi}^2}/{r}$ and $\displaystyle{GM}/{r^2}$). Therefore,
in computations of the trajectories of particles
whose dynamics are determined by the comparable
actions of the centrifugal and inward gravitational
forces, high-order methods require less computational
resources than first-order methods (here, we are referring
to methods for the solution of an arbitrary system
of ordinary differential equations, and not specialized
simplex methods adapted for specific problems). We
investigated how the requirements for the numerical
method change in computations of particle trajectories
influenced by additional forces.

\begin{table}
\caption{\label{tab:methodsErr} Errors in the values of variables in the system (\ref{eq:system}), found using two different methods: an explicit fourth-order
Runge–Kutta method and the first-order semi-implicit method, relative to the reference solution}
\begin{center}
\begin{tabular}{ |c|c|c|c| } 
\hline
Model—Drag coefficient & Weidenschilling (1977) & Henderson (1976) \\
\hline
\multirow{3}{*}{Toy Model~1} & $\Delta r/r = 1.3453 \times 10^{-5}$ & $\Delta r/r = 7.837 \times 10^{-6}$ \\ 
& $\Delta v_r/v_r = 1.3718 \times 10^{-3}$ & $\Delta v_r/v_r = 8.3969 \times 10^{-4}$ \\ 
& $\Delta v_\varphi/v_\varphi = 4.297 \times 10^{-6}$ & $\Delta v_\varphi/v_\varphi =3.358 \times 10^{-6}$\\ 
\hline
\multirow{3}{*}{Toy Model~2} & $\Delta r/r = 4.1573 \times 10^{-5}$ & $\Delta r/r = 1.0109 \times 10^{-5}$ \\ 
& $\Delta v_r/v_r = 9.4828 \times 10^{-4}$ & $\Delta v_r/v_r = 4.1845 \times 10^{-4}$\\ 
& $\Delta v_\varphi/v_\varphi = 1.4979 \times 10^{-5}$ & $\Delta v_\varphi/v_\varphi = 1.8898 \times 10^{-5}$\\ 
\hline
\end{tabular}
\end{center}
\end{table}

For this, we computed Toy Model 1 and Toy
Model 2 using an explicit fourth-order Runge–Kutta
method with the same step, $\tau=0.01\Omega^{-1}(r=1$~au). We determined the values of $r$, $v_r$ and $v_{\varphi}$ at the time
when the particle crossed $r=1$~au, and further compared these values with those computed using the
semi-implicit first-order scheme. The results of this
comparison are presented in Table~{\ref{tab:methodsErr}}. In both models
and with both drag coefficients, the accuracy of the
computations for the first-order scheme differs from
that for the fourth-order scheme by less than 1\%, even though the Stokes number for the large particles
exceeds $10^2$ in both models. Among the three variables
of the system, the maximum relative errors are displayed
by $v_r$. The computations for Toy Model 2 using
the different methods and drag coefficients differ only
insignificantly. However, in Toy Model 1, where the
standard astrophysical drag coefficient has a discontinuity,
the use of the Henderson coefficient makes it
possible to obtain more similar solutions using the
first- and fourth-order methods. 

In order to investigate the differences in the accuracy
of the solutions obtained using the first- and
fourth-order methods in more detail, we first compared
the solutions of the system (\ref{eq:system}) without drag
force. This system of three equations has two first integrals,
which express the conservation of angular
momentum and energy:
\begin{equation}
\label{eq:systemC1}
C_1=r v_{\varphi},
\end{equation}

\begin{equation}
\label{eq:systemC2}    
C_2=\displaystyle \frac{v_r^2}{2} + \frac{v_{\varphi}^2}{2}-\frac{GM}{r}. 
\end{equation}

\begin{figure}
    \centering
    \includegraphics[width=170mm]{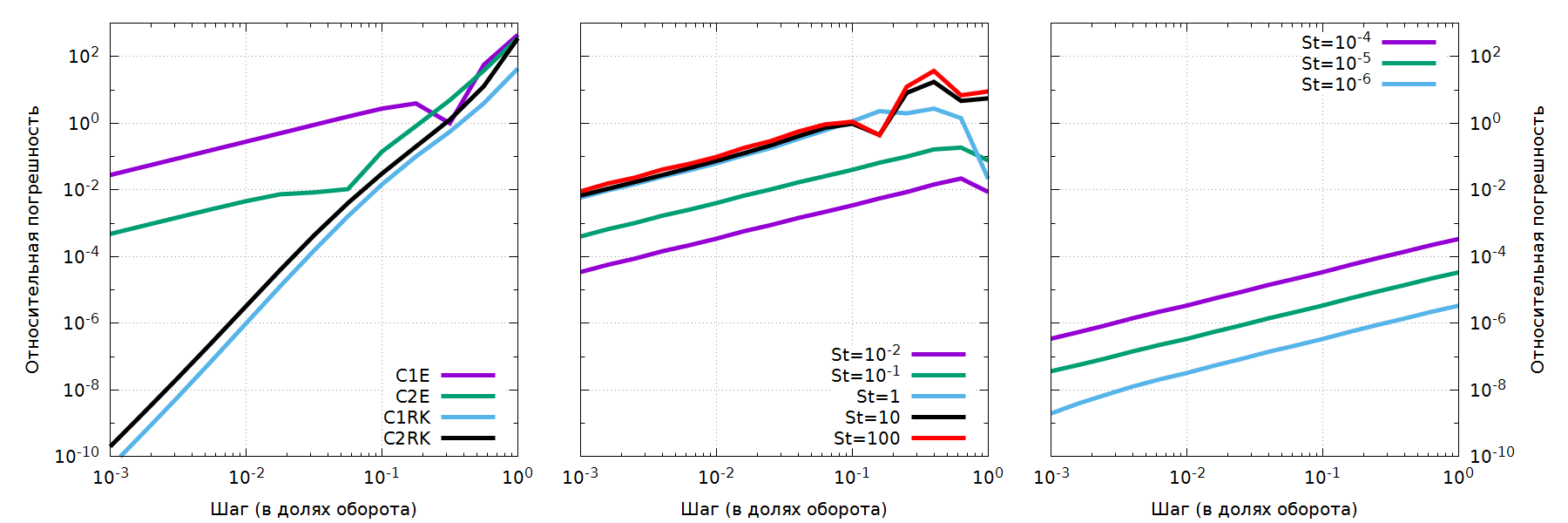}
    \caption{The left panel shows the relative error in the solution of the system without friction after 40 orbital periods [the integrals (\ref{eq:systemC1}), (\ref{eq:systemC2})], obtained using the first- and fourth-order methods in the time step, the central and right panels the errors in $v_r$ in the
solution of the system with drag using the first-order semi-implicit method for various Stokes numbers. The integration time was
two orbital periods for the central panel and 40 orbital periods for the right panel.}
    \label{fig:accuracy}
\end{figure}

At the initial time, we placed the particle at the radius $r_0=10$~au and specified its radial velocity to be close to
zero and its azimuthal velocity to be the Keplerian
value. We computed the particle’s trajectory over a
time $t=40 \times 2 \pi \Omega^{-1}(r_0)$ (40 orbits around the central
body) using an explicit Euler method (into which the
semi-implicit, first-order scheme with zero drag is
degenerate) and the explicit Runge–Kutta method. At
the final time, we calculated the values of $C_1$ and $C_2$ and compared them to the initial values of these integrals.
The dependence of the relative errors in $C_1$ and $C_2$ on the time step $\tau$ re presented in the left panel of
Fig.~{\ref{fig:accuracy}}. Both schemes give the stated order of approximation
in the solution, with the accuracy of the computation
of the angular momentum in the first-order
scheme being almost a factor of 100 worse than the
accuracy of the computation of the energy. Computing $C_1$ with an accuracy of up to 1\% using the first-order scheme requires more than 1000 steps in one orbit of
the central body, whereas 10 steps per orbit is sufficient
for the Runge–Kutta method.

We also studied how many steps per orbit are
required by the first-order scheme in order to obtain
an accuracy of 1\% when drag force is added to the
system (\ref{eq:system}). We solved the system (\ref{eq:system}), where the
components of the drag force have the form
$\left(\displaystyle\frac{(v_r-u_r)\Omega(r)}{\rm{St}},\displaystyle\frac{(v_{\varphi}-u_{\varphi})\Omega(r)}{\rm{St}}\right)$ with the constant value
of $\rm{St}$. For $\rm{St}=10^{-6}, \quad 10^{-5}, \quad 10^{-4}$ we carried out the
integration over a time $t=40 \times 2 \pi \Omega^{-1}(r_0)$, and for $\rm{St}=10^{-2}, \quad 10^{-1}, \quad 1, \quad 10, \quad 100$ over the shorter time $t=2 \times 2 \pi \Omega^{-1}(r_0)$. We used the solution of this same problem
obtained using the fourth-order method with a time
resolution of $10^{7}$ steps per orbit $t= 2 \pi \Omega^{-1}(r_0)$, as a reference.
The relative errors in $v_r$, computed in the semi-implicit,
first-order scheme are presented in the central
and right panels of Fig.~\ref{fig:accuracy}. For all values of $\rm{St}$ the
semi-implicit scheme demonstrates first-order in its
approximation, but the actual accuracy attained in the
solution depends on $\rm{St}$. In the computations presented
in the right panel, $\rm{St} \ll 1$, and the dynamics of the particle
are determined by the gas drag, not by the balance
between centrifugal and gravitational forces. In these
computations, the lower the value of $\rm{St}$, the higher the
accuracy of the numerical solution. In the computations
with $\rm{St} \ge 1$, presented in the central panel, the
dynamics of the particle are determined by the balance
between the centrifugal and gravitational forces, and
the actual accuracy in $v_r$ does not depend on $\rm{St}$ ,and is
close to the accuracy with which $C_1$ is calculated in
the computations without friction using the same time
steps. 

On the whole, the data presented in Fig.~{\ref{fig:accuracy}} and
Table~\ref{tab:methodsErr} show the expediency of using the first-order
scheme for simulations of the dynamics of particles in
a disk at orbital radii greater than 1 au.

\section{CONCLUSION}
\label{sec:resume}

The sizes of dust grains in circumstellar and protoplanetary
disks vary from submicron to several centimeters,
and planetesimals have sizes of hundreds of
kilometers. Therefore, the exchange of momentum
between solid bodies and gas in disks occurs in a variety
of regimes, determined by the sizes and velocities
of the solid bodies: Epstein, Stokes, and Newton. We
have compared different drag coefficients encompassing
these regimes. We have shown that the standard
astrophysical drag coefficient (\ref{eq:CD4}) is a discontinuous
function for Mach numbers corresponding to the
motion of the grains relative to the gas $\rm{Ma}>1/9$, but
is continuous when $\rm{Ma}< 1/9$. Application of the standard
astrophysical coefficient in simulations of circumstellar
disks with $\rm{Ma}<1/9$ is justified by the speed
of the computations. When $\rm{Ma}>1/9$, it is recommended
to use the Henderson drag coefficient (\ref{eq:HendersonSub})--(\ref{eq:HendersonTrans}), which is valid and continuous when $\rm{Ma}<6$, $\rm{Re}<3 \times 10^{5}$. Each computation of the Henderson
coefficient requires of about 100 arithmetic operations.

In addition, the need to compute the dynamics of
bodies of different sizes in the same way imposes high
demands on numerical methods used for this purpose.
We have shown that our semi-implicit, first-order
approximation scheme, in which the interphase interactions
are linearized and the relative velocity is calculated
implicitly, while the stopping time and other
forces, such as the pressure gradient and gravitation
are calculated explicitly, preserve the asymptotic
behavior of the solutions of problems with intense
interphase interactions. This means that the scheme is
suitable for computing the dynamics of both small
bodies that are strongly coupled to the gas and larger
bodies for which the drag force depends non-linearly
on the relative velocity between the gas and bodies, including the influence of the dispersed phase on the
gas dynamics.

\appendix

\section{APPROACH TO DESCRIBING
THE DYNAMICS OF DUST
IN A CIRCUMSTELLAR DISK}

The suitability of a hydrodynamical approach to
describing a medium is usually determined using two
conditions:
\begin{itemize}
    \item The mean free path of the particles making up
the medium should be much shorter than the length
scale of the system. We will take the disk scale height $H$ as a length scale of the system. 
\end{itemize}
Further, some element of the medium is distinguished
whose size is much less than the length scale
of the medium, but much larger than the mean free
path of the particles. In our case, this is the size of a
computational grid cell, $\delta R$.
\begin{itemize}
    \item The number of particles in a given volume of the
medium $\delta R$ should be sufficiently large.
\end{itemize}

The first condition is required so that the loss/gain
of particles by an element of the medium as a result of
chaotic motions of the particles does not appreciably
influence the mean characteristics of the element. The
second condition is required for computations using
the particle distribution function in the six-dimensional
space (coordinates and velocities) of the mean
characteristics, such as the mean density in a cell, the
mean velocity, and the mean energy.

There is no doubt that these conditions are satisfied
by the gas in a circumstellar disk. However, we must
investigate this question for the dust component. We
estimated the mean free path of a solid particle assuming that the dust subdisk consists of monodisperse,
spherical particles. In this case,
\begin{equation}
\label{eq:freepass}  
    \lambda_{\rm d}=\displaystyle\frac{1}{\sqrt{2} \pi a^2 n_{\rm d}}=\displaystyle \frac{4a \rho_{\rm s}}{2 \sqrt{2} \rho_{\rm d}},
\end{equation}
since, by virtue of their monodisperse nature,
\begin{equation}
    n_{\rm d}(a)=\displaystyle\frac{3\rho_{\rm d}}{4\pi\rho_{\rm s}a^3}.
\end{equation}

We assumed that the volume density of the dust in the
equatorial plane of the disk was given by
\begin{equation}
    \rho_{\rm d}=\frac{\Sigma_{\rm d}}{H\sqrt{\pi}}, 
\end{equation}
where the height of the disk is
\begin{equation}
    H=\sqrt{\frac{2k_{\rm B}T}{\mu_{\rm H}}\frac{r^3}{GM}}
\end{equation}
Let the surface density of the dust and the temperature
be power-law functions of the radius, $\Sigma_{\rm d}(r)=\Sigma_{0}(r/1\,{\textrm{AU}})^p$, $T=T_{0}(r/1\,{\textrm{AU}})^q$. 
Adopting
the characteristic values for a circumstellar disk  $\Sigma_{0}=1$~g~cm$^{-2}$, $T_0$=300~K, $\mu_{\rm H}=2.3$, $M=1M_{\odot}$, $p=-1, q=-0.5$, we obtain
\begin{equation}
\label{eq:dustmidplane}
    \rho_{\rm d}=7.6\cdot 10^{-13} (r/1\,{\textrm{AU}})^{-2.25} \quad{\textrm{g~cm}}^{-3}.
\end{equation}
It follows from (\ref{eq:freepass}) and (\ref{eq:dustmidplane})that grains smaller than
1 cm in the outer parts of a disk have mean free paths
much shorter than $\delta R$. It is clear that the dust subdisk
can be modeled as a continuous medium in this case.
On the other hand, these estimates indicate that $\lambda_{\rm d}$ is close to $H$ in inner regions of the disk. In this case,
instead of the mean velocity over the volume, we must
consider the velocity of the solid phase as a vector plus
a dispersion. A possible development of the model
could be the use of the Vlasov–Boltzmann equation \cite{1MNRAS} or moments of the Boltzmann equation \cite{VorobyovTheis2006}. The
solution of the Vlasov–Boltzmann equation in
Lagrange coordinates implies the use of discrete particles
when modeling the dynamics of the solid phase \cite{BaiStone2010ApJS,Zhu2014,YangJohansen2016}. Solution for the moment Boltzmann equation
can be found in an Euler approach using grid
methods \cite{VorobyovTheis2008}. 

The moment Boltzmann equation take into
account the possible anisotropic behaviour of the
velocity dispersion of the medium, but computing the
components of the velocity dispersion requires the
introduction of additional equations. When solving
the Vlasov–Boltzmann equation using a particle
method, the velocity dispersion varies in a self-consistent
way \cite{BoothClarke2016}, but the difficulty is correctly computing
the collision integral. In gas dynamics, this integral is
zero for frequent elastic collisions. This may not be the
case for the grains.

We have concluded that dust in the main part of the
disk can be represented as a continuous medium, while there may be regimes in the central region of the
disk in which solid particles have a velocity dispersion,
based on estimates of the mean free paths of particles,
neglecting their interactions with the gas. However,
the same conclusion is supported by estimates of the
spatial scale on which exchanges of momentum
between the gas and dust occur:
\begin{equation}
    \lambda_{\rm stop}=\|u-v\|t_{\rm stop}, 
\end{equation}
where $t_{\rm stop}={\rm St}/\Omega_{\rm K}$. According to \cite{Birnstiel2016} Eq.(8), the
azimuthal velocity of a particle can be estimated as
\begin{equation}
    u=v_{\rm K}\left(1-\frac{1}{1+{\rm St}^2}\eta\right).
\end{equation}
Here, $\eta$ characterizes the degree of deviation of the
velocity of the gas from the Keplerian value, and, for
typical disk parameters, $\eta \sim 0.01$. We then find that
\begin{equation}
    \lambda_{\rm stop}=r{\rm St}\left(1-\frac{1}{1+{\rm St}^2}\eta\right) \sim  r{\rm St}.
\end{equation}

It is clear that, when $St < 0.01$, the velocities of grains
within a single cell will differ little, while they will differ
appreciably when $St \sim 1$. The simulation results \cite{VorobyovEtAl2017} indicate that dust that has grown is located in the inner
part of the disk.

\section{MODEL OF A STATIONARY GASEOUS DISK}

\begin{figure}
    \centering
    \includegraphics[width=140mm]{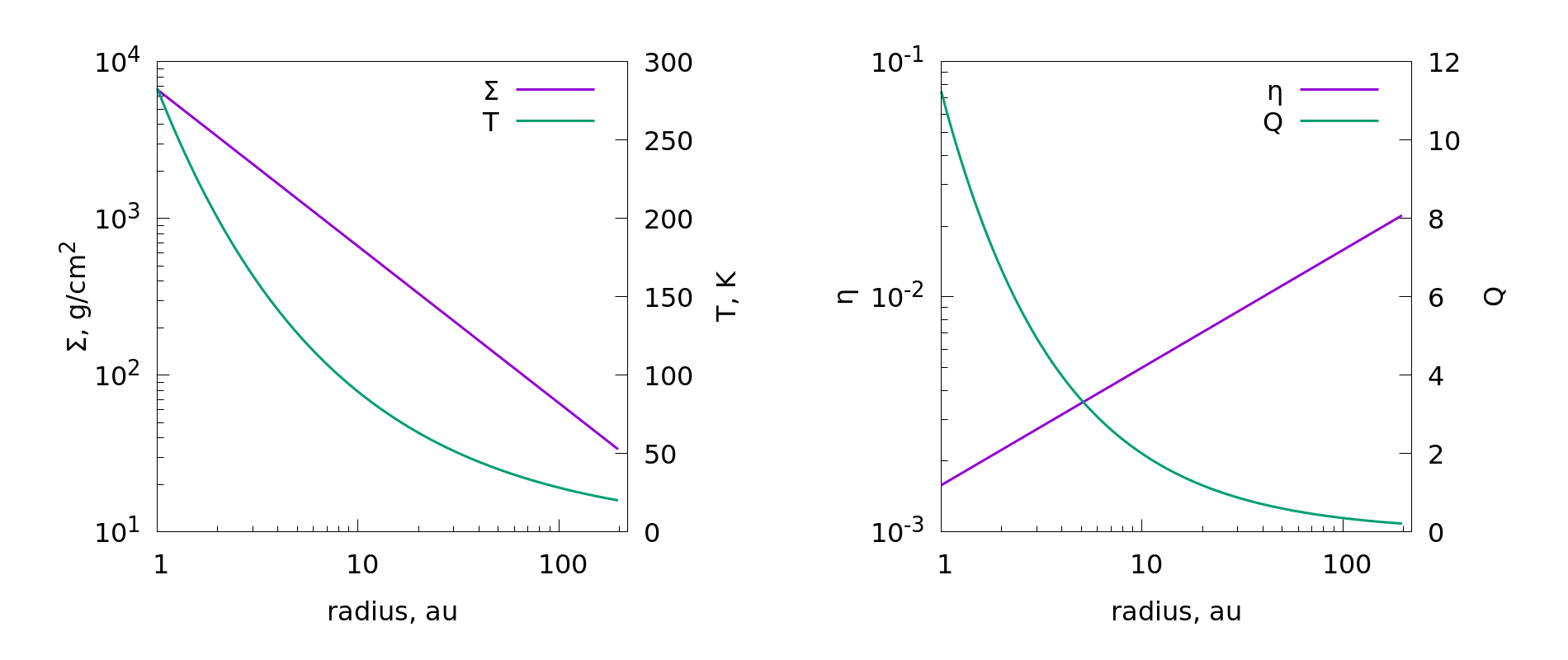}
    \caption{(Left panel) Radial distributions of the surface density and temperature in a circumstellar disk for Toy Models 1 and 2.
(Right panel) Radial distribution of $\eta$ and Toomre parameter for Toy Models 1 and 2.}
    \label{fig:ToySigmaT}
\end{figure}

We calculated the radial distributions of parameters
of an axially symmetric, gaseous disk with inner
boundary $R_{\rm min}$ and outer boundary  $R_{\rm max}$ with mass $M_{\rm disk}$ around a star with mass $M$ using the model of \cite{RiceEtAl2004}, which is based on the following assumptions:
\begin{itemize}
\item the surface density of the matter in the disk displays
a power-law dependence on the radius with
power-law index $\xi$:
\begin{equation}
\Sigma(r)=\Sigma_0\left(\frac{r}{R_0}\right)^{-\xi},
\end{equation}
where $\Sigma_0 R_0^{\xi}$ is given by

\begin{equation}
 \label{eq:Sigma0}
\int_{0}^{2 \pi} \int_{R_{\rm min}}^{R_{\rm max}} \Sigma(r) r \,dr\,d\varphi =M_{\rm{disk}},
\end{equation}

\end{itemize}

\begin{itemize}
\item the density and temperature of the gas in the disk
are such that the Toomre parameter varies as
\begin{equation}
\label{eq:Toomre}
Q(r)=\frac{c_{s}(r)\Omega(r)}{\pi G \Sigma(r)} = 2\left(\frac{R_{\rm{max}}}{r}\right)^{0.75}, 
\end{equation}

\item the gaseous disk is in equilibrium, so that
\begin{equation}
\label{eq:uDisc}
u_r=0, \quad u_{\varphi}(r)=\sqrt[]{\displaystyle\frac{GM}{r}+\frac{r}{\rho_{\rm g}(r)}\frac{d p(r)}{d r}},
\end{equation}
\item the height $H$ and radius $r$ of the disk are related as
\begin{equation}
\displaystyle\frac{c_{\rm s}(r)}{V_{\rm K}(r)}=\frac{H(r)}{r},  
\label{Eq:HightRadiusRef}
\end{equation}
\item the gas density at each radius does not vary in the
vertical direction, so
\begin{equation}
\rho_{\rm g}(r)=\displaystyle\frac{\Sigma(r)}{H(r)}.
\end{equation}
\end{itemize}

We assumed $\xi=1$ and used the above assumptions
to find the distributions $c_{\rm s}(r)$, $\rho_{\rm g}(r)$, $u_{\varphi}(r)$ in the disk,
which are required for the solution of (\ref{eq:system}).

It follows from (\ref{eq:Sigma0}) that
\begin{equation}
\Sigma_0=\frac{M_{\rm{disk}}(2-\xi)}{2\pi R_0^{\xi}(R_{\rm{max}}^{2-\xi}-R_{\rm{min}}^{2-\xi})}.
\end{equation}

It follows from (\ref{eq:Toomre}) that
\begin{equation}
  c_s(r)=\frac{Q(r) \pi G \Sigma(r)}{\Omega(r)}=\frac{\pi r Q(r) G \Sigma(r) }{V_{\rm K}(r)},  
\end{equation}
and therefore
\begin{equation}
\label{eq:c2Vk}
    \frac{c_s(r)}{V_{\rm K}(r)}=
    \frac{\pi Q(r) G \Sigma(r) r}{V_{\rm K}^2}=
    \frac{\pi Q(r) \Sigma(r) r^2}{M}.
\end{equation}
We then obtain from (\ref{eq:c2Vk}) and (\ref{Eq:HightRadiusRef})
\begin{equation}
\label{eq:H}
    H(r)=\displaystyle\frac{\pi Q(r) \Sigma(r) r^3}{M},
\end{equation}
whence
\begin{equation}
 \rho_{\rm g}(r)=\displaystyle\frac{\Sigma(r)}{H(r)}=\frac{M}{\pi Q(r) r^3}.   
\end{equation}
Note that, by virtue of (\ref{eq:Toomre}) 
\begin{equation}
\label{eq:dlnrho}
  \rho_{\rm g}(r)=\rho_{\rm g 0} \displaystyle \left(\frac{r}{R_0}\right)^{-2.25}.  
\end{equation}


We took $u_\varphi$ to be related to $V_{\rm K}$ as
\begin{equation}
\label{eq:uVarphiDisc}
  u_\varphi^2=V_{\rm K}^2+\left(\frac{d\ln\rho_{\rm g}}{d\ln r}\right)c_s^2=V_{\rm K}^2(1-\eta),  
\end{equation}
where
\begin{equation}
   \eta=\frac{c_s^2}{V_{\rm K}^2}\left|\frac{d\ln\rho_{\rm g}}{d\ln r}\right|=2.25 \frac{c_s^2}{V_{\rm K}^2}=\frac{2.25 \pi^2 Q^2(r) \Sigma^2(r) r^4}{M^2}, 
\end{equation}
so that
\begin{equation}
u_\varphi= V_{\rm K}\,\sqrt[]{1-\frac{2.25 \pi^2 Q^2(r) \Sigma^2(r)r^4}{M^2}} = V_{\rm K}\,\sqrt[]{1-\frac{2.25 \Sigma^2(r)}{\rho^2_{\rm g}(r)r^2}} \end{equation}

The distribution of the surface density and temperature
for the disk in Toy Models 1 and 2 is shown in
Fig.~\ref{fig:ToySigmaT}.



\section{ACKNOWLEDGMENTS}

We thank V.N. Snytnikov and M.I. Osadchii for productive
discussions of our results.

\section{FUNDING}
This work was founded by the Russian Science Foundation
(grant 17-12-01168).

\newpage


\end{document}